\documentclass[useAMS,usenatbib]{mn2e}

\usepackage{times}
\usepackage{graphicx}
\usepackage{subfigure}

\newcommand{\gsim}{\mathrel{\hbox{\rlap{\lower.55ex \hbox {$\sim$}}
                   \kern-.3em \raise.4ex \hbox{$>$}}}}
\newcommand{\lsim}{\mathrel{\hbox{\rlap{\lower.55ex \hbox {$\sim$}}
                   \kern-.3em \raise.4ex \hbox{$<$}}}}

\title[Radiative feedback and the IMF]{The importance of radiative feedback for the stellar initial mass function}
\author[M.R. Bate]{Matthew R. Bate\thanks{E-mail:
mbate@astro.ex.ac.uk}\\ School of Physics, University of Exeter, Stocker
Road, Exeter EX4 4QL}

\date{Accepted by MNRAS}
\begin{document}
\maketitle
\begin{abstract}
We investigate the effect of radiative feedback on the star formation process using radiation hydrodynamical simulations.  We repeat the previous hydrodynamical star cluster formation simulations of Bate et al., and Bate \& Bonnell, but we use a realistic gas equation of state and radiative transfer in the flux-limited diffusion approximation rather than the original barotropic equation of state.  

Whereas star formation in the barotropic simulations continued unabated until the simulations stopped, we find that radiative feedback, even from low-mass stars, essentially terminates the production of new objects within low-mass dense molecular cloud cores after roughly one local dynamical time.  Radiative feedback also dramatically decreases the propensity of massive circumstellar discs to fragment and inhibits fragmentation of other dense gas (e.g. filaments) close to existing protostellar objects.  These two effects decrease the numbers of protostars formed by a factor of $\approx 4$ compared with the original hydrodynamical simulations using the barotropic equation of state.  In particular, whereas the original simulations produced more brown dwarfs than stars, the radiative feedback results in a ratio of stars to brown dwarfs of approximately 5:1, in much better agreement with observations.  Most importantly, we find that although the characteristic stellar mass in the original calculations scaled linearly with the initial mean Jeans mass in the clouds, when radiative feedback is included the characteristic stellar mass is indistinguishable for the two calculations, regardless of the initial Jeans mass of the clouds.  We thus propose that the reason the observed initial mass function appears to be universal in the local Universe is due to self-regulation of the star formation process by radiative feedback.  We present an analytic argument showing how a characteristic mass may be derived that is relatively independent of initial conditions such as the cloud's density.
\end{abstract}
\begin{keywords}
stars: formation, stars: low-mass, brown dwarfs, stars: luminosity function, mass function, methods: numerical, radiative transfer, hydrodynamics.
\end{keywords}

\section{Introduction}
\label{introduction}

Understanding the origin of the stellar initial mass function (IMF) is one of the fundamental goals of a complete theory of star formation.  Observationally, much has been accomplished in the five decades since \citet{Salpeter1955} published his seminal paper on the form of the IMF.  Salpeter's determination of the high-mass end of the IMF has become widely accepted, but the behaviour of the IMF below 1 M$_\odot$ has been more accurately characterised, with the turnover into the brown dwarf regime being the subject of many recent investigations.  The general form of the IMF in the solar neighbourhood is now known down to $\approx 0.03$~M$_\odot$ \citep{Kroupa2001,Chabrier2003}.  However, despite this progress  in determining the form of the IMF, there is still no standard model for the origin of the IMF or on how it should depend on environment.  In fact, much of the difficulty in determining which processes are responsible for the origin of the IMF can be attributed to the fact that the IMF doesn't seem to vary strongly with environment \citep[for recent reviews see][]{Kroupa2002, Chabrier2003}.  Galactic studies and those extending to the Magellanic Clouds have repeatedly failed to find any systematically robust and statistically significant variation from the general form of the IMF.  There is evidence from observations of the Arches Cluster and the stars orbiting the supermassive black hole, Sgr A$^{*}$, that the IMF near the Galactic Centre may be biased in favour of massive stars.  However, the apparently top-heavy IMF in the Arches Cluster \citep{Figeretal1999, Stolteetal2002} may be due to dynamical evolution rather than being primordial in origin \citep{Portegiesetal2002, Kimetal2006, Portegiesetal2007}, and the apparent deficit of low-mass stars surrounding Sgr A$^{*}$ is only indirectly inferred from the lack of X-ray emission \citep{NaySun2005}.  This has led to the use of the term `universal IMF' when describing present-day star formation.  Only for the first, zero metallicity, stars in the Universe is there general agreement that the IMF should differ from that found locally, producing substantially more massive stars \citep*[e.g.,][]{BroCopLar1999, AbeBryNor2000}.  However, this has yet to be confirmed observationally. 

Many theories have been proposed for the origin of the IMF.  These fall into 
four main classes.  The IMF may originate from fragmentation, whether it be turbulent fragmentation \citep*{Larson1992, Henriksen1986, Henriksen1991, Elmegreen1997, Elmegreen1999, Elmegreen2000, PadNorJon1997, PadNor2002}, gravitational fragmentation \citep{Larson1973,ElmMat1983, Zinnecker1984, YosSai1985}, or domain packing \citep{Richtler1994}, with the fragmentation subject to an opacity limit which sets a minimum stellar mass \citep{Hoyle1953, Gaustad1963, Yoneyama1972, SucShc1976, LowLyn1976, Rees1976, Silk1977a, Silk1977b, MasInu1999}.  It may depend on feedback processes such as winds and outflows \citep{Shuetal1988, Silk1995, AdaFat1996} or heating of accretion discs around massive black holes \citep{Nayakshin2006}.  It may originate from competitive accretion of fragments (\citealt{Hoyle1953, Larson1978, Zinnecker1982, Bonnelletal1997, Bonnelletal2001a}a; \citealt{Bonnelletal2001b}b; \citealt*{KleBurBat1998}; \citealt{Myers2000}). Or it may be due to coalesence or collisional build-up \citep*{SilTak1979, PumSca1983, BonBatZin1998, BonBat2002}.  In reality, all of these processes are likely to play some role.  The main questions to answer are which process, if any, dominates the origin the the IMF, and how does the IMF vary with environment?

The most fundamental parameter of the IMF is its characteristic mass, $\sim 0.3 {\rm M}_\odot$.  A lower limit to the mass of a `star' is provided by the opacity limit for fragmentation of a few Jupiter masses for typical molecular gas \citep{Hoyle1953, LowLyn1976, Rees1976, WhiSta2006}.  All objects must have masses greater than this minimum mass.  However, the peak of the IMF occurs at masses roughly two orders of magnitude greater than this minimum mass.  Several theories for the IMF link the characteristic mass to the typical Jeans mass in the progenitor molecular cloud.  This may be the thermal Jeans mass \citep[e.g.][]{Larson1992}, a magnetic critical mass, or a turbulent Jeans mass \citep{Silk1995}.  A Jeans mass origin for the characteristic stellar mass has been supported by some hydrodynamical calculations of the fragmentation of clumpy and turbulent molecular clouds.  In these calculations, it was found that the mean mass of the protostars was similar to the initial mean Jeans mass of the cloud (\citealt{KleBurBat1998,KleBur2000,KleBur2001,Klessen2001}; \citealt*{BatBonBro2003}) and that variations in the initial Jeans mass led to corresponding variations in the characteristic mass of the IMF \citep{BatBon2005,Bate2005}.  Another quite different model of the IMF proposes that the IMF orginates from the mass distribution of dense cores in turbulent molecular clouds \citep{PadNor2002}.  In this model, the characteristic mass of the IMF depends on both the initial Jeans mass and the Mach number of the turbulence.  However, as pointed out by \citet{AdaFat1996}, there is no unique Jeans mass in a molecular cloud.  \citet{Larson1985,Larson2005} proposes that the appropriate Jeans mass may be that at which gas and grains couple thermally and dust cooling takes over from atomic line cooling in molecular clouds.  \citet*{WhiBofFra1998} also link the characteristic mass of the IMF with this gas-grain coupling point.  Hydrodynamical simulations using an equation of state inspired by Larson's models and varying it to give a different characteristic Jeans mass do produce corresponding changes in the characteristic mass of the IMF (\citealt{Jappsenetal2005}; \citealt*{BonClaBat2006}).  Recently, \citet*{ElmKleWil2008} have argued that the characteristic Jeans mass at which the gas-dust coupling occurs is relatively independent of environmental quantities such as density, temperature, metallicity, and the radiation field, perhaps explaining the apparent universality of the characteristic mass of the IMF.

In this paper, we investigate the effect of radiative feedback on the star formation process.  We repeat the hydrodynamical calculations of star cluster formation \citet{BatBonBro2003} and \citet{BatBon2005} that resolved fragmentation down to the opacity limit, circumstellar discs, and binary stars (hereafter BBB2003 and BB2005, respectively).  The initial conditions for these two calculations were identical except that the latter cloud was denser than the former, lowering the initial mean Jeans mass by a factor of 3.  BB2005 showed that the characteristic masses of the two IMFs produced depended linearly on the initial Jeans mass, also differing by a factor of 3.  Here, we repeat the original calculations, but instead of using a barotropic equation of state to model the thermodynamic behaviour of the gas, we use a realistic gas equation of state and include radative transfer in the flux-limited diffusion approximation.  We find that {\em the inclusion of radiative feedback from the forming protostars substantially weakens the dependence of the characteristic mass of the IMF on the initial Jeans mass in the progenitor molecular cloud.  Thus, we propose that star formation regulates itself via radiative feedback to provide the IMF with a characteristic mass that is usually only weakly dependent on environment.}

The paper is structured as follows. In Section \ref{method}, we briefly describe the numerical method and the initial conditions for the simulations.  We present the results from our radiation hydrodynamical simulations in Section \ref{results}.  In Section \ref{discussion}, we compare our results with previous simulations and with observations, and we discuss the implications of our results for the origin of the characteristic mass of the IMF. Our conclusions are given in Section \ref{conclusions}.

\section{Computational method}
\label{method}

The calculations presented here were performed 
using a three-dimensional SPH code based on a 
version originally developed by Benz 
\citep{Benz1990, Benzetal1990} but substantially
modified as described in \citet{BatBonPri1995},
\citet*{WhiBatMon2005} and \citet{WhiBat2006} and 
parallelised by M.\ Bate using OpenMP.

Gravitational forces between particles and a particle's 
nearest neighbours are calculated using a binary tree.  
The smoothing lengths of particles are variable in 
time and space, subject to the constraint that the number 
of neighbours for each particle must remain approximately 
constant at $N_{\rm neigh}=50$.  The SPH equations are 
integrated using a second-order Runge-Kutta-Fehlberg 
integrator with individual time steps for each particle
\citep{BatBonPri1995}.
We use the standard form of artificial viscosity 
\citep{MonGin1983, Monaghan1992} with strength 
parameters $\alpha_{\rm_v}=1$ and $\beta_{\rm v}=2$.

\subsection{Equation of state and radiative transfer}

The original hydrodynamical calculations of BBB2003 and
BB2005 modelled the thermal behaviour of the gas 
without performing radiative transfer using a barotropic equation of state.

In this paper, we use an ideal gas equation of state $p= \rho T \cal{R}/\mu$, where $\rho$ is the density, $T$ is the gas temperature, $\cal{R}$ is the gas constant, 
and $\mu$ is the mean molecular
mass.  The equation of state takes into account the translational,
rotational, and vibrational degrees of freedom of molecular hydrogen
(assuming an equilibrium mix of para- and ortho-hydrogen; see
\citealt{BlaBod1975}).  It also includes the dissociation of molecular
hydrogen, and the ionisations of hydrogen and helium.  
The hydrogen and helium mass fractions are $X=0.70$ and 
$Y=0.28$, respectively.
The contribution of metals to the equation of state is neglected.  Further
details on the implementation of the equation of state can be found
in \citet{WhiBat2006}.

Two temperature (gas and radiation) radiative transfer in the flux-limited
diffusion approximation is implemented as described by \citet{WhiBatMon2005}
and \citet{WhiBat2006}.  Energy is generated when work is done on the gas
or radiation fields, radiation is transported via flux-limited diffusion, 
and energy is transferred between the gas and radiation fields depending 
on their relative temperatures, the gas density, and the gas opacity.  We use 
interpolation from the opacity tables of \citet{PolMcKChr1985} to provide 
the interstellar grain opacities for solar metallicity molecular gas, whilst at 
higher temperatures when the grains have been destroyed we use the tables
of \citet{Alexander1975} (the IVa King model) to provide the gas opacities
(for further details, see \citealt{WhiBat2006}).  The only change from 
\citet{WhiBat2006} is to do with the boundary condition.  The clouds modelled
here have free boundaries.  To provide a boundary condition for the 
radiative transfer, all particles with densities less than $10^{-21}$ g~cm$^{-3}$
have their gas and radiation temperatures set to the initial values of 10 K.  This gas 
is $2-3$ orders of magnitude less dense that the initial clouds (see below) and,
thus, these boundary particles surround the region of interest in which the
star cluster forms.

\subsection{Sink particles}
\label{sinks}

\begin{table*}
\begin{tabular}{lcccccccccc}\hline
Calculation & Initial Gas & Initial  & Jeans & Mach & Accretion & No. Stars & No. Brown  & Mass of Stars and  & Mean & Median \\
 & Mass  & Radius & Mass & Number & Radii & Formed & Dwarfs Formed & Brown Dwarfs & Mass & Mass \\
 & M$_\odot$ & pc & M$_\odot$ & & AU & & & M$_\odot$ & M$_\odot$ & M$_\odot$\\ \hline
BBB2003 & 50.0 & 0.188 & 1 & 6.4 & 5 & $\geq$23 & $\leq$27 & 5.89 & 0.118 & 0.070 \\
BBB2003 RT5 &  & &  &  & 5 & $\geq$10 & $\leq$5 & 7.09 & 0.473 & 0.22~~ \\
BBB2003 RT0.5 &  & & & & 0.5 & $\geq$11 & $\leq$2 & 6.76 & 0.520 & 0.31~~ \\ \hline
BB2005 & 50.0 & 0.090 & 1/3 & 9.2 & 5 & $\geq$19 & $\leq$60 & 7.92 & 0.100 & 0.023 \\
BB2005 RT0.5 & & & &  & 0.5 & $\geq$14 & $\leq$3 & 7.57 & 0.446 & 0.35~~  \\ \hline
\end{tabular}
\caption{\label{table1} The initial conditions and the statistical properties of the stars and brown dwarfs formed in the original BBB2003 and BB2005 calculations and the new versions of those calculations using radiation hydrodynamics that we present here.  Calculations BBB2003 RT0.5 and BB2005 RT0.5 use radiation hydrodynamics and sink particles with accretion radii of $r_{\rm acc}=0.5$ AU.  Calculation BBB2003 RT5 is identical to BBB2003 RT0.5 except that the sink particles have $r_{\rm acc}=5$ AU.  The initial conditions for the two different types of calculation are identical except that the BB2005-type initial cloud has a smaller radius giving a density 9 times higher and a mean thermal Jeans mass a factor of 3 lower.  All calculations were run for 1.40 initial cloud free-fall times.  Brown dwarfs are defined as having final masses less than 0.075 M$_\odot$.  The numbers of stars (brown dwarfs) are lower (upper) limits because some of the brown dwarfs were still accreting when the calculations were stopped.  Using radiation hydrodynamics dramatically reduces the numbers of objects formed, particularly brown dwarfs.  Furthermore, whereas using a barotropic equation of state led to the median mass scaling linearly with the mean thermal Jeans mass of the cloud, radiation hydrodynamics removes the dependence of the median stellar mass on the initial Jeans mass.}
\end{table*}

In the original hydrodynamical calculations of BBB2003 and
BB2005, gas collapsed isothermally until a density of $10^{-13}$~g~cm$^{-3}$ 
beyond which point the gas temperature increased with density as 
$\rho^{2/5}$.  This resulted in the formation of pressure-supported 
fragments with initial masses of a few Jupiter-masses in collapsing regions.
As each fragment accreted its central density and temperature increased, 
resulting in smaller and smaller timesteps. When the central density of a fragment
exceeded $10^{-11}$~g~cm$^{-3}$, it was replaced by a sink particle.

In this paper, the evolution a collapsing region of gas is similar except that the
temperature is calculated self-consistently using radiative transfer and the realistic
equation of state and we
follow fragments to higher densities before inserting a sink particle.  Collapsing
gas that exceeds $\sim 10^{-13}$~g~cm$^{-3}$ becomes optically thick and
traps radiation, thus heating up and forming a pressure-supported fragment
known as the `first hydrostatic core' \citep{Larson1969}.  This core continues
to accrete mass and its central density and temperature rise until 
molecular hydrogen begins to dissociate at a temperature of $\approx 2000$~K
and density of $\sim 10^{-7}$~g~cm$^{-3}$.  The dissociation of molecular hydrogen 
requires energy, initiating a second dynamical collapse within the first
core.  When the density exceeds $\sim 10^{-3}$ g~cm$^{-3}$ the gas is 
atomic and a second, `stellar', core is formed.  We follow the gas through the
entire first core phase and through most of the second collapse phase, inserting
a sink particle when the density exceeds $10^{-5}$ g~cm$^{-3}$ 
(in terms of the real star formation process, this is just a few {\em days} before 
the stellar core is formed).

In the original barotropic calculations, a sink particle was formed by 
replacing the SPH gas particles contained within $r_{\rm acc}=5$ AU 
of the densest gas particle 
by a point mass with the same mass and momentum.  Any gas that 
later fell within this radius was accreted by the point mass 
if it was bound and its specific angular momentum was less than 
that required to form a circular orbit at radius $r_{\rm acc}$ 
from the sink particle.  Thus, gaseous discs around sink 
particles could only be resolved if they had radii $\gsim 10$ AU.
Sink particles interacted with the gas only via gravity and accretion.
The angular momentum accreted by a sink particle was recorded
but played no further role in the calculation.
The gravitational acceleration between two 
sink particles was Newtonian for $r\geq 4$ AU, but was softened within this 
radius using spline softening \citep{Benz1990}.  The maximum acceleration 
occurred at a distance of $\approx 1$ AU; therefore, this was the
minimum separation that a binary could have even if, in reality,
the binary's orbit would have been hardened.  

In this paper, our default is to use accretion radii of only $r_{\rm acc}=0.5$ AU and we
do not use any gravitational softening between two sink particles.  This allows
us to resolve smaller discs and closer binaries than in the earlier 
hydrodynamical simulations.  We also performed a calculation based on the BBB2003 initial
conditions with 
accretion radii of $r_{\rm acc}=5$ AU to see what difference this made to the results.
Sink particles are permitted to merge if they
pass within 0.02 AU of each other (i.e., $\approx 4$~R$_\odot$).
This radius was chosen because recently formed protostars are
thought to have relatively large radii \citep[e.g.,][]{Larson1969}.
No mergers occurred in any of the calculations reported here.

Finally, we emphasise that sink particles only interact with the rest of the
simulation via gravity and by accreting gas.  In particular, for the radiative hydrodynamical
calculations presented here {\em there is no 
radiative feedback from the sink particles into the rest of the simulation}.
Neglecting the intrinsic luminosity of an accreting low-mass protostar is 
reasonable since the accretion luminosity overwhelms the intrinsic luminosity
of the object even for a very low accretion rate.  However, since gas is not
modelled within $r_{\rm acc}$ of each protostar due to the sink particle approximation, 
a substantial fraction of the
accretion luminosity is also neglected (from the gravitational energy liberated 
during the inspiral of gas from $r_{\rm acc}$ to the stellar surface).  Thus, we 
emphasise that {\em these calculations give a lower limit on the effects of radiative feedback on the star formation process}.  
We perform one of the calculations with accretion radii of both 5 AU and 0.5 AU
specifically to investigate the effect of this approximation on the results.

\subsection{Initial conditions}
\label{initialconditions}

The initial conditions for the two calculations are identical to those 
of BBB2003 and BB2005 and are summarised in Table \ref{table1}.  Each spherical cloud contains 
50 M$_\odot$ of molecular gas.  The radii of the two clouds are 0.188 pc and 0.090 pc, respectively, so that the latter cloud has 9 times the density of the former
(densities of $1.2\times 10^{-19}$ and $1.1\times 10^{-18}$ g~cm$^{-3}$, respectively).
At the initial temperature of 10~K, the two clouds have mean thermal 
Jeans masses of 1 and 1/3 M$_\odot$, respectively.  
Although each cloud is uniform in density initially, an initial supersonic 
`turbulent' velocity field is imposed on each of them in the same manner
as \citet*{OstStoGam2001}.  
The divergence-free random Gaussian velocity field has 
a power spectrum $P(k) \propto k^{-4}$, where $k$ is the wavenumber.  
In three dimensions, this results in a
velocity dispersion that varies with distance, $\lambda$, 
as $\sigma(\lambda) \propto \lambda^{1/2}$ in agreement with the 
observed Larson scaling relations for molecular clouds 
\citep{Larson1981}.
The velocity field was generated on a $128^3$ uniform grid and the
velocities of the particles were interpolated from the grid.  The same 
velocity field is used for each of the two clouds, but the normalisation
differs and is set so that the the kinetic energy 
of the turbulence equals the magnitude of the gravitational potential 
energy of each cloud.
Thus, the initial root-mean-square (rms) Mach number of the turbulence 
was ${\cal M}=6.4$ in BBB2003 and ${\cal M}=9.2$ in BB2005.
The initial free-fall times of the two clouds are $t_{\rm ff}=6.0\times 10^{12}$~s or 
$1.90\times 10^5$ years and $t_{\rm ff}=2.0\times 10^{12}$~s or 
$6.34\times 10^4$ years, respectively.

In fact, the early evolution of each of the clouds was not repeated using 
radiation hydrodynamics since the gas remains essentially isothermal during the
early evolution with radiative transfer, but the radiation hydrodynamical calculations 
are approximately an order of magnitude more computationally expensive 
than the barotropic calculations.  The radiation 
hydrodynamical calculations were instead begun from dump files from 
the original calculations just before the density exceeded 
$1\times 10^{-16}$ g~cm$^{-3}$ and $9 \times 10^{-16}$ g~cm$^{-3}$, respectively.

\subsection{Resolution}

The local Jeans mass must be resolved throughout the calculations to model fragmentation correctly 
(\citealt{BatBur1997, Trueloveetal1997, Whitworth1998, Bossetal2000}; \citealt*{HubGooWhi2006}).  This requires $\gsim 1.5 N_{\rm neigh}$ SPH particles per Jeans mass; $N_{\rm neigh}$ is insufficient (BBB2003).  
The original calculations used $3.5 \times 10^6$ particles to model the 
50-M$_\odot$ clouds and resolve the Jeans mass down to its minimum 
value of $\approx 0.0011$ M$_\odot$ (1.1 M$_{\rm J}$) at the maximum 
density during the isothermal phase of the collapse, 
$\rho_{\rm crit} = 10^{-13}$ g~cm$^{-3}$.  Using radiation hydrodynamics results in
temperatures at a given density no less than those given by the original 
barotropic equation of state and, thus, the Jeans mass is also resolved in
the calculations presented here.

Each of the two calculations with sink particle accretion radii of $r_{\rm acc}= 0.5$ AU 
required the equivalent of approximately 40,000 CPU hours on a 16-processor 1.65GHz IBM p570 
compute node of the United Kingdom Astrophysical Fluids Facility (UKAFF).  One of
the two calculations was completed on the University of Exeter Supercomputer, an SGI Altix ICE 8200.  
The $r_{\rm acc}=5$ AU calculation ran approximately 20 times faster.

\begin{figure*}
\centering \vspace{-0.0cm}
    \includegraphics[width=7.3cm]{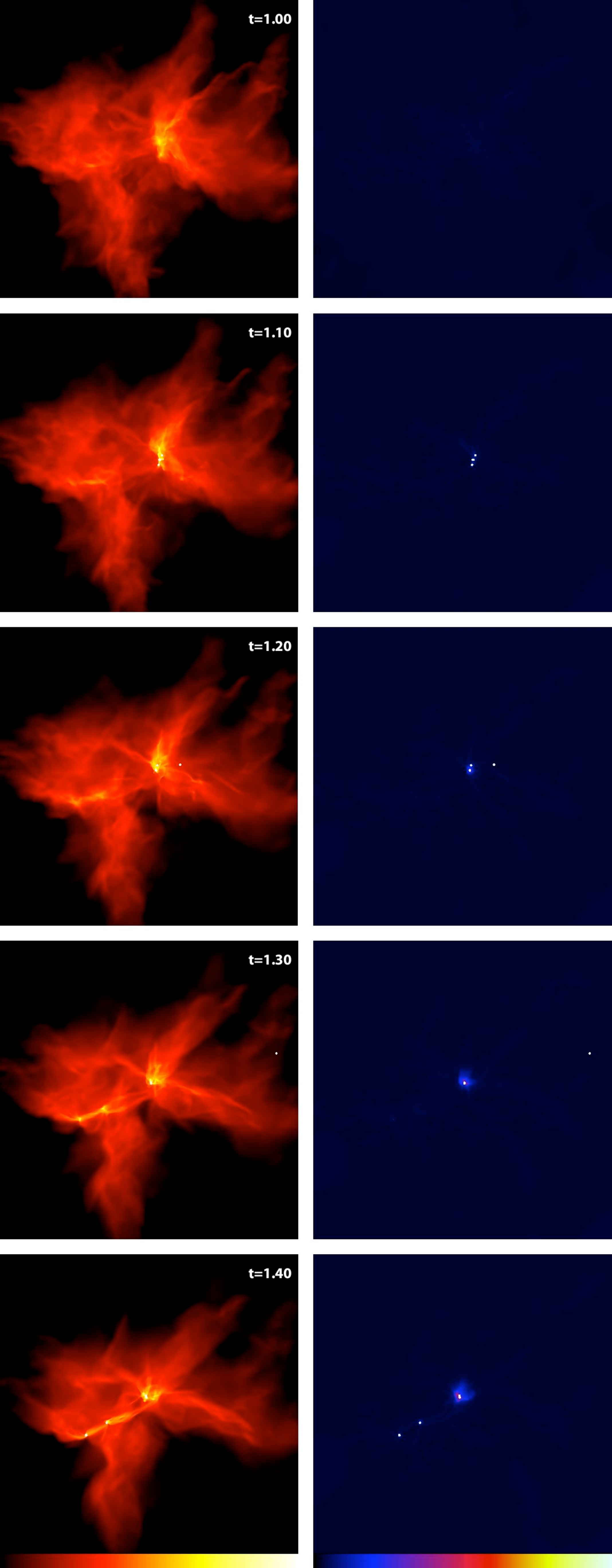}\hspace{1.5cm}
    \includegraphics[width=7.3cm]{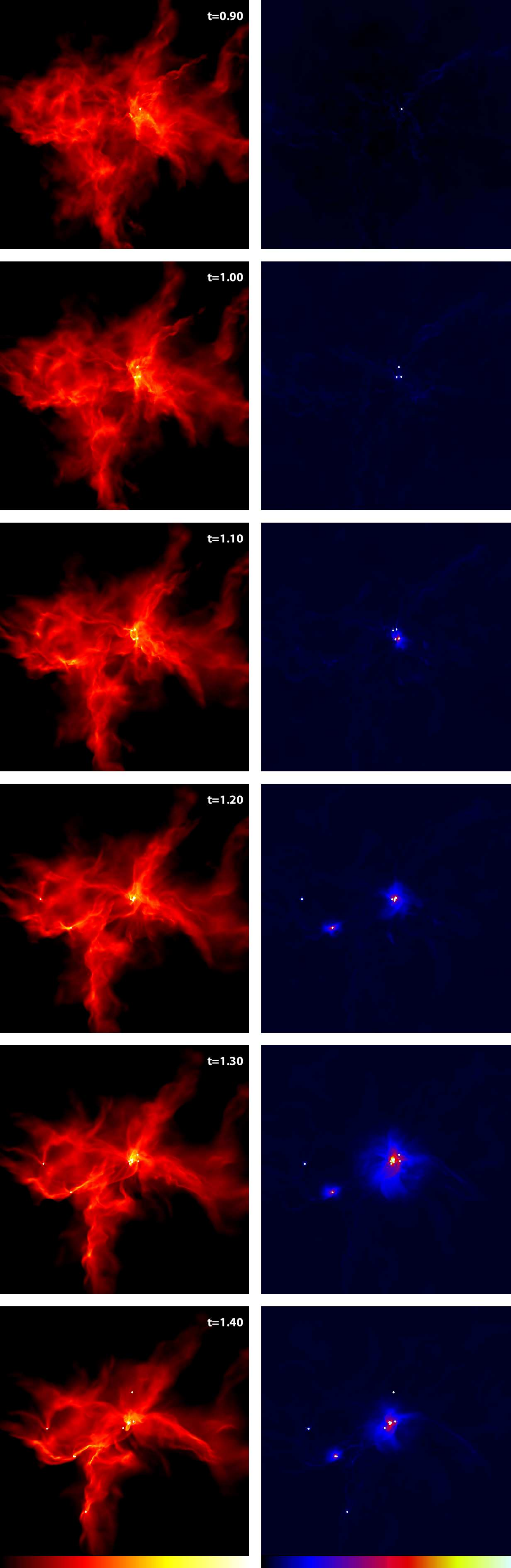}
\caption{Global evolution of the BBB2003 RT0.5 (left) and BB2005 RT0.5 (right) calculations.  In each case, the left (red-yellow-white) panels show the logarithm of column density, $N$, in g~cm$^{-2}$ and the scales cover $-1.7<\log N<0.5$ (left) and $-1.1<\log N<2.1$ (right).  The right (blue-red-yellow-white) panels show the logarithm of mass weighted temperature, $T$, in K with the scales covering $9-100$ K.  The lefthand panels are 0.4 pc (82,400 AU) across while the righthand panels are 0.192 pc (39,600 AU) across.  Time is given in units of the initial free-fall time of $1.90\times 10^5$ yr (left) and $6.34 \times 10^4$ yr (right).}
\label{global}
\end{figure*}

\begin{figure*}
\centering \vspace{-0.0cm}
    \includegraphics[width=15.8cm]{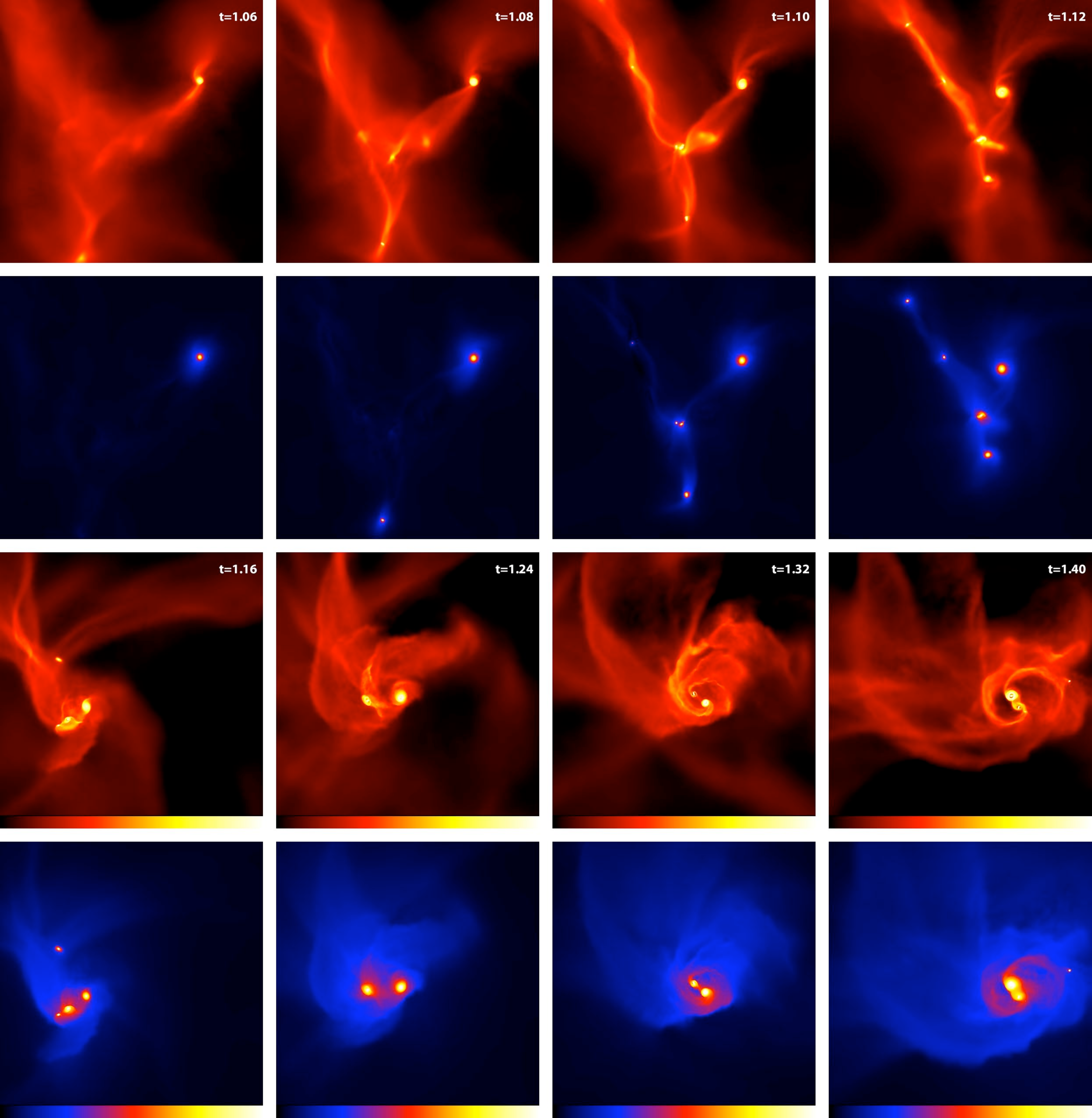}
\caption{The star formation in the main dense core of the BBB2003 RT0.5 calculation. The first object forms at $t=1.042~t_{\rm ff}$. Large gaseous filaments collapse to form single objects and multiple systems. These objects fall together to form a small group. Radiative feedback from the accreting protostars heats the gas in the dense core.  Each panel is 0.025 pc (5150 au) across. These may be compared to the equivalent figures in the original BBB2003 paper.  Time is given in units of the initial free-fall time of $1.90\times 10^{5}$ yr. The red-yellow-white panels show the logarithm of column density, N, through the cloud, with the scale covering $-0.5 < \log N < 2.5$ with N measured in g~cm$^{-2}$.  The blue-red-yellow-white panels show the logarithm of mass weighted temperature, $T$, through the cloud with the scale covering $9-300$~K. }
\label{BBB2003core}
\end{figure*}

\begin{figure*}
\centering \vspace{-0.0cm}
    \includegraphics[width=7.6cm]{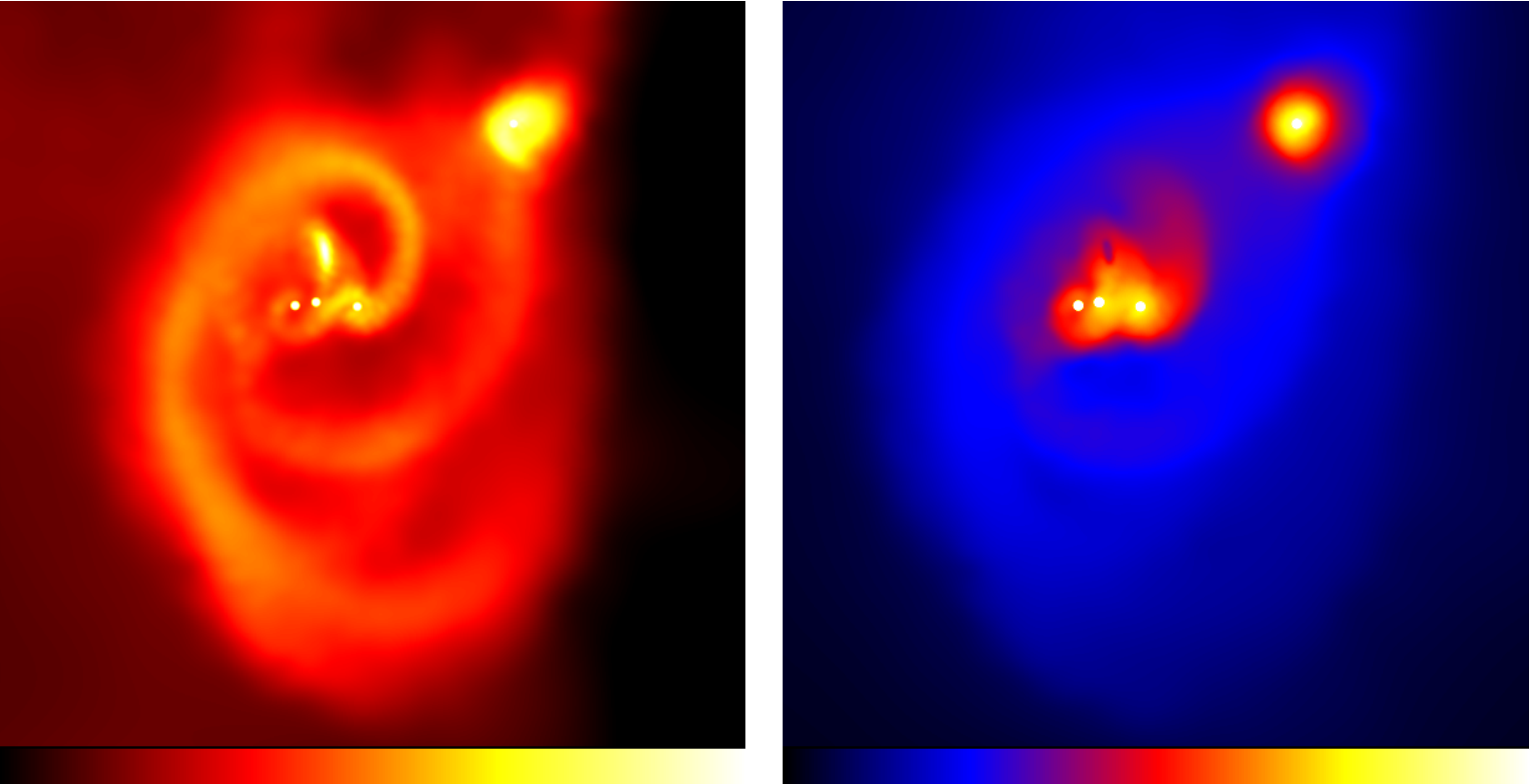}\hspace{1cm}
    \includegraphics[width=7.6cm]{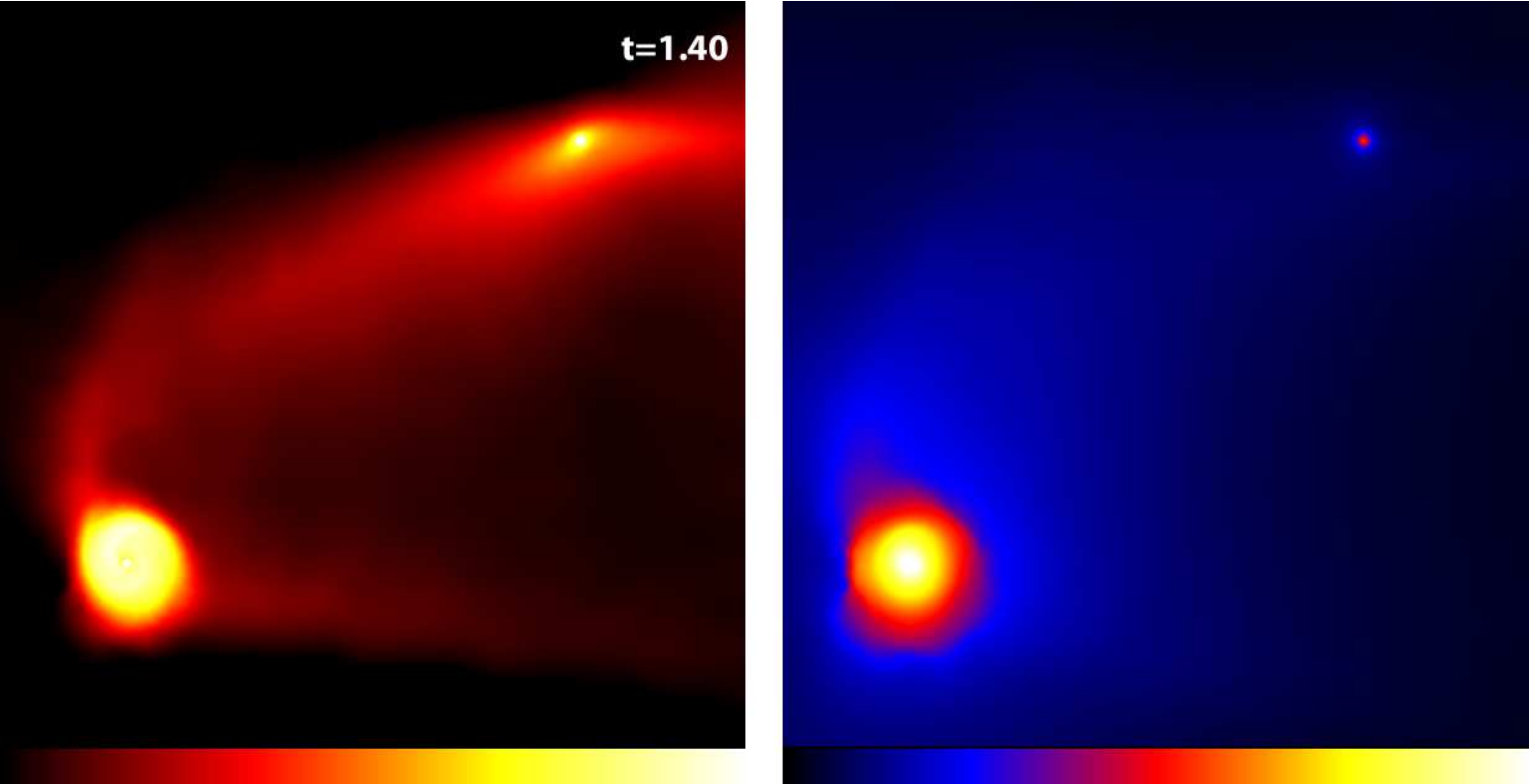}
\caption{The star formation in the second (left) and third (right) dense cores at the end of the BBB2003 RT0.5 calculation. The left two panels are 600 AU across, while the right two panels are 1000 AU across. These may be compared to the equivalent figures in the original BBB2003 paper.  Time is given in units of the initial free-fall time of $1.90\times 10^{5}$ yr. The red-yellow-white panels show the logarithm of column density, N, through the cloud, with the scale covering $0.0 < \log N < 2.5$ with N measured in g~cm$^{-2}$.  The blue-red-yellow-white panels show the logarithm of mass weighted temperature, $T$, through the cloud with the scale covering $9-300$~K. At the end of the calculation, the object in the top-right corner of the panels depicting the third core had not yet undergone the second collapse and been replaced by a sink particle.}
\label{BBB2003core23}
\end{figure*}

\section{Results}
\label{results}

Results from three new calculations are presented in this paper.  All were performed using the radiation hydrodynamical SPH code described above.  Calculations BBB2003 RT0.5 and BB2005 RT0.5 are identical the hydrodynamical calculations presented in BBB2003 and BB2005, respectively, except that they were performed using radiation hydrodynamics and sink particles have accretion radii of 0.5 AU, ten times smaller than in the original calculations.  Calculation BBB2003 RT5 is identical to BBB2003 RT0.5 except that the sink particles have accretion radii of 5 AU, the same size as in the original BBB2003 calculation.  This last calculation enables us to determine how sensitive the results are to the size of the accretion radii.  This is even more important when using radiation hydrodynamics than for a barotropic equation of state because sink particles themselves do not emit radiation -- only the gas emits radiation (see Section \ref{sinks}).  Thus, using smaller accretion radii allows more accretion luminosity to be released by the protostars into the calculations, leading to hotter gas and potentially affecting the pattern of fragmentation.  Each of the radiation hydrodynamical simulations is followed to $1.40~t_{\rm ff}$, the same as the original BBB2003 and BB2005 calculations.

 \subsection{BBB2003 initial conditions}
 \label{BBB2003text}

We begin by presenting the results from the two BBB2003-type calculations using radiation hydrodynamics.  As mentioned in Section \ref{initialconditions}, the radiation hydrodynamical calculations were not re-run from the initial conditions, but were started from the last dump file from the original BBB2003 before the density exceeded $10^{-16}$ g~cm$^{-3}$.  Before this point the initial `turbulent' velocity field had generated density structure in the gas, some of which was collected into dense cores which had begun to collapse.  Those readers interested in this early phase should refer to BBB2003 for figures and further details.

The BBB2003 radiation hydrodynamical calculations were started from $t=0.976~t_{\rm ff}$ (in the original BBB2003 calculation the first sink particle was inserted at $t=1.037~t_{\rm ff}$, some $1.2\times 10^4$ years later).  Using radiation hydrodynamics, the first sink particle is inserted at $t=1.042~t_{\rm ff}$.  The slightly later time is primarily because in the original calculation sink particles were inserted when the density exceeded $10^{-11}$ g~cm$^{-3}$ (when the fragment was in the `first hydrostatic core' stage of evolution) whereas with the radiation hydrodynamics we do not insert sink particles until halfway through the second collapse phase at a density of $10^{-5}$ g~cm$^{-3}$ (see \ref{sinks} for further details).

\subsubsection{Sink particles with $r_{\rm acc}=0.5$ AU}

\begin{figure}
\centering
    \includegraphics[width=8.4cm]{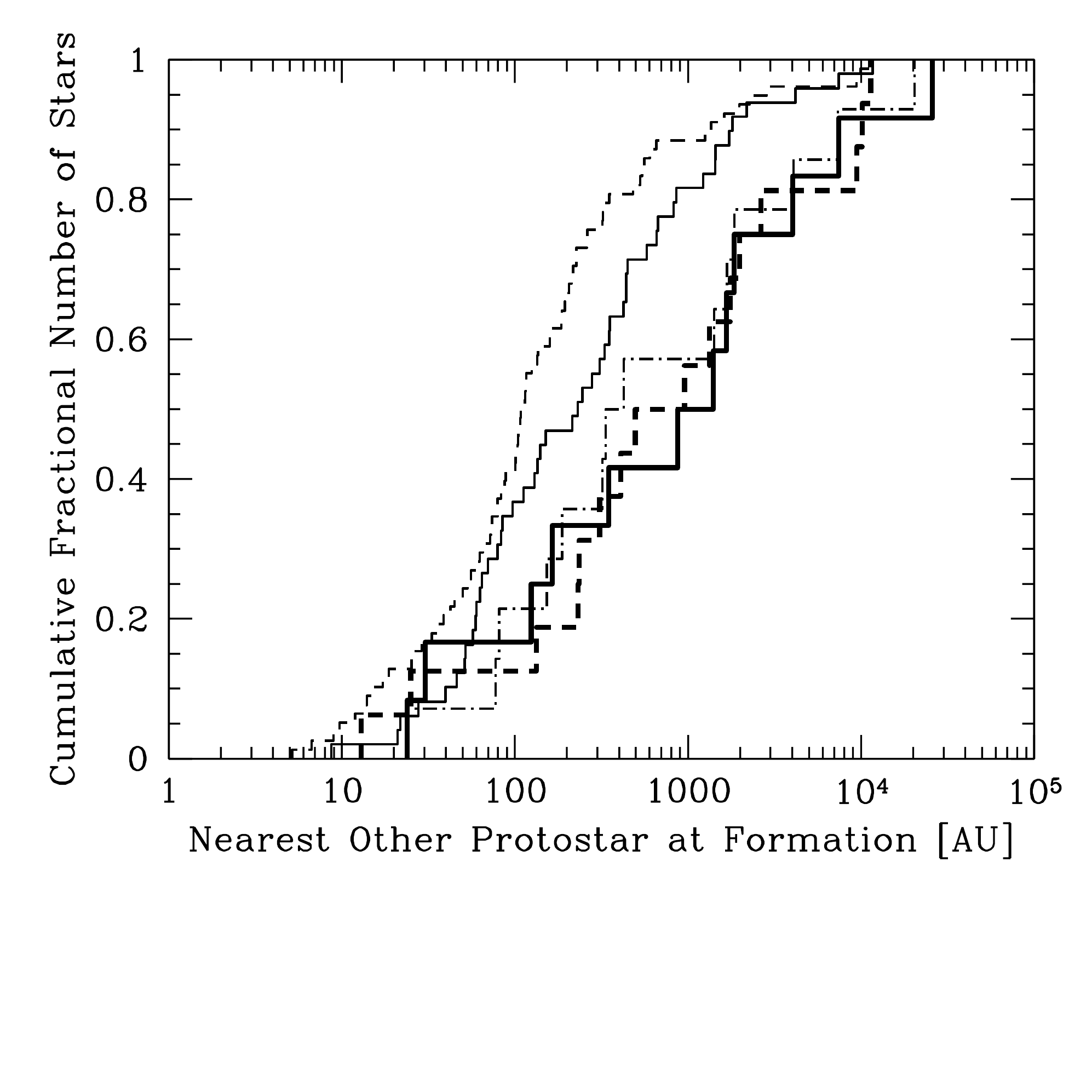}\vspace{-1.5cm}
\caption{The cumulative distributions of the distance between a new protostar and its closest other protostar (excluding the first protostar to form in each calculation) for each of the 5 calculations discussed in this paper.  The distributions from the previously published BBB2003 and BB2005 calculations using a barotropic equation of state are given by the thin solid line and thin dashed line, respectively.  The radiation hydrodynamical calculations presented here are BBB2003 RT0.5 (thick solid line), BB2005 RT0.5 (thick dashed line), and the large accretion radius BBB2003 RT5 calculation (thin dot-dashed line).  It is clear that including radiative feedback increases the typical distance between a new protostar and its closest companion.  For the BB2005-type initial conditions, 3/4 of the objects formed within 300 AU of another protostar using a barotropic equation of state, whereas with radiative feedback this fraction is reduced to less than 1/3.  For the BBB2003-type initial conditions, more than 70\% of the objects formed within 500 AU of another protostar using a barotropic equation of state, whereas with radiative feedback and small accretion radii this percentage is reduced to just over 40\%.}
\label{sepform}
\end{figure}

\begin{figure}
\centering
    \includegraphics[width=7.5cm]{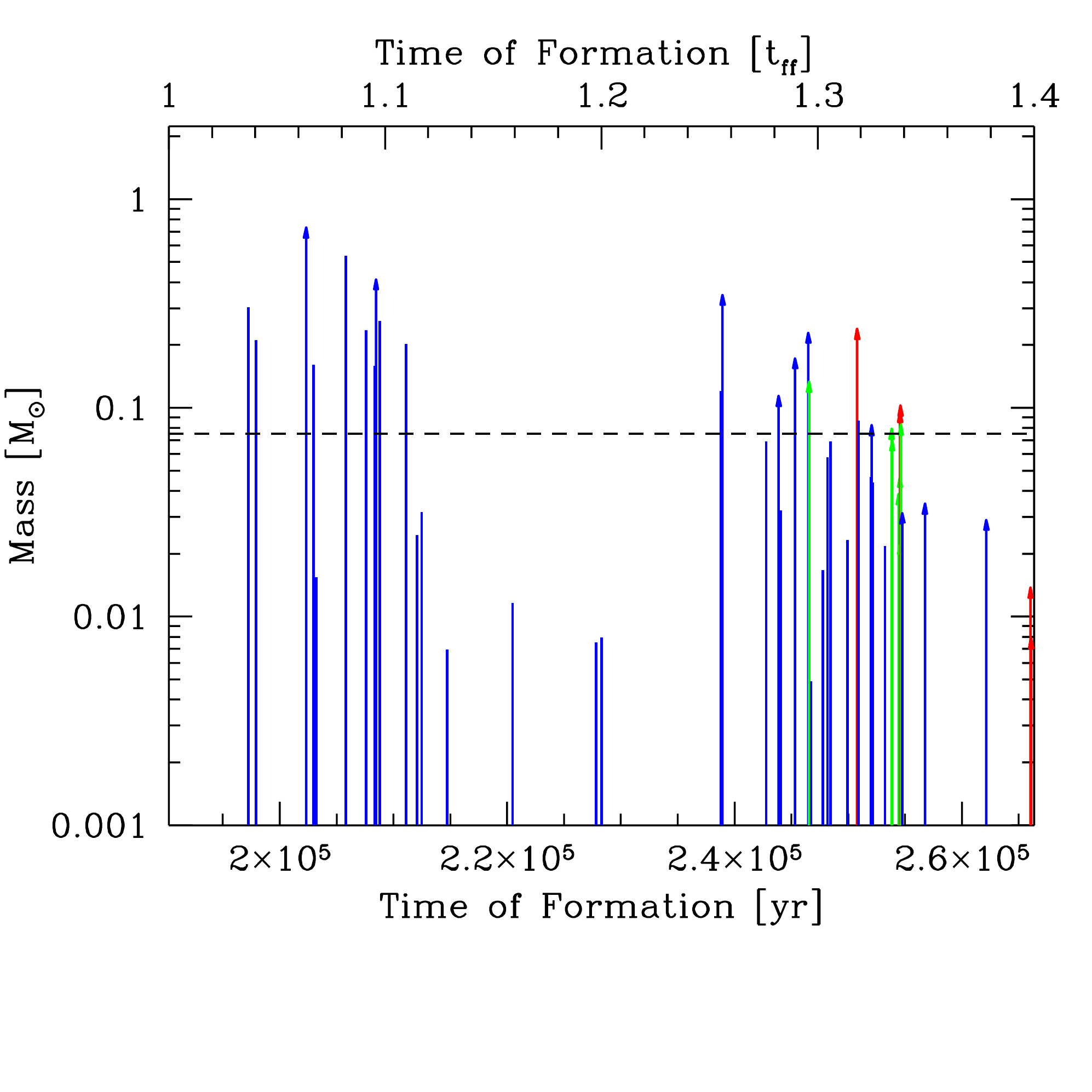}\vspace{-1.0cm}
    \includegraphics[width=7.5cm]{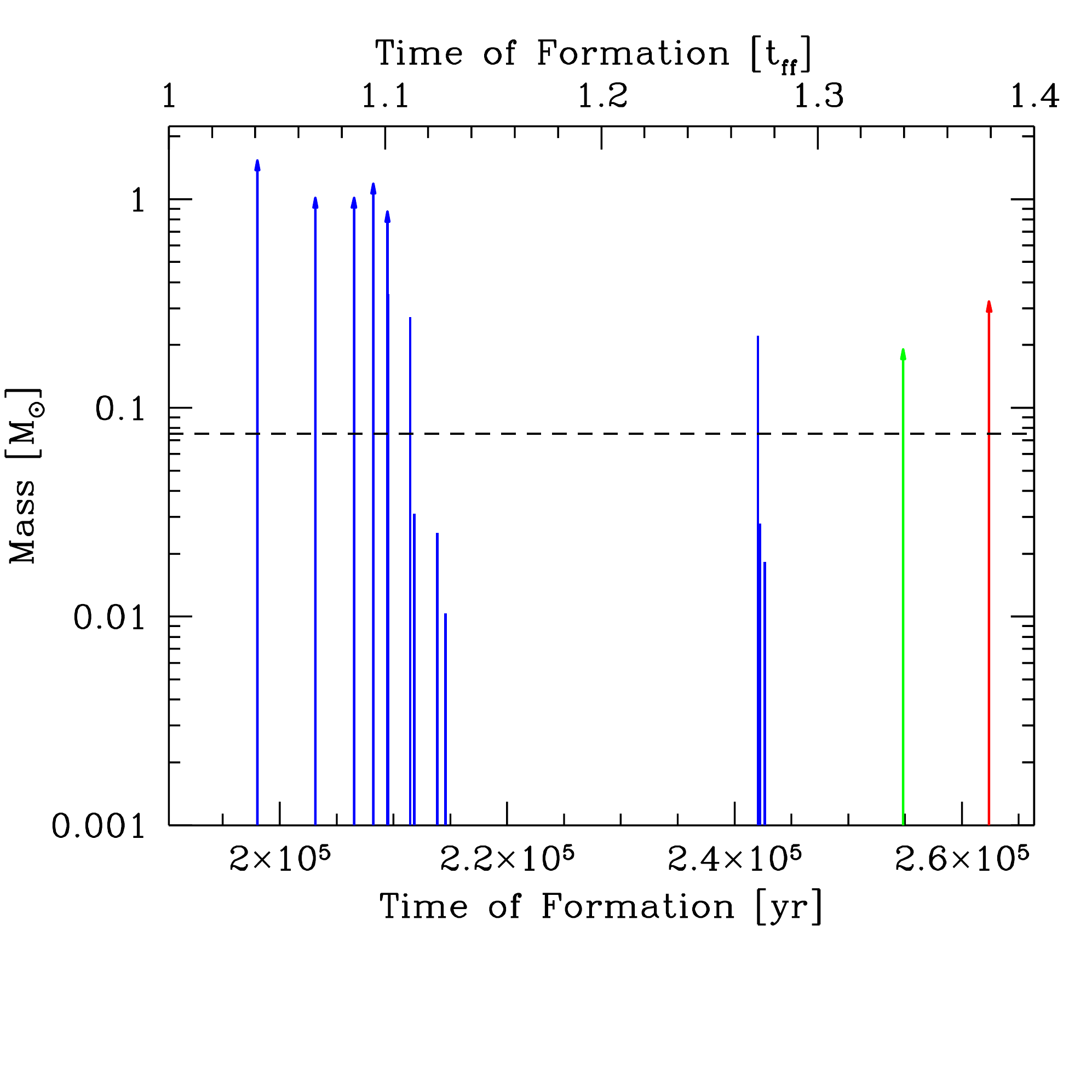}\vspace{-1.0cm}
    \includegraphics[width=7.5cm]{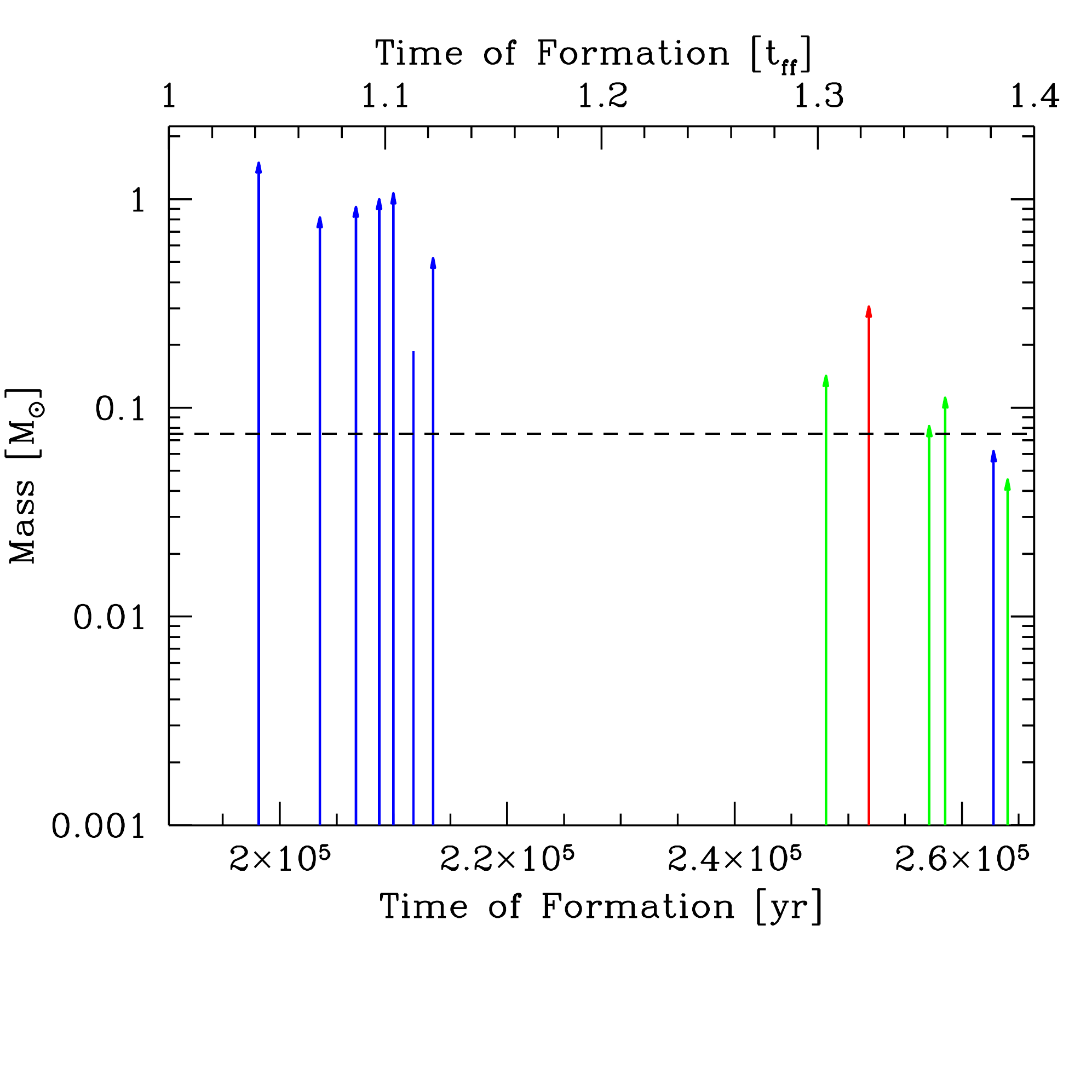}\vspace{-1.2cm}
\caption{Time of formation and mass of each star and brown dwarf at the end of original BBB2003 calculation (top) and the radiation hydrodynamical calculations BBB2003 RT5 (centre) and BBB2003 RT0.5 (bottom).  Radiative feedback dramatically decreases the number of objects formed, primarily by suppressing the continued star formation in the main dense core that occurs in the original calculation.  
Objects that form in the main dense core are denoted by blue lines.  Objects that form in the lower-mass second and third dense cores are denoted by green and red lines, respectively.  Objects that are still accreting significantly at the end of the calculation are represented with vertical arrows.  The horizontal dashed line marks the star/brown dwarf boundary.  Time is measured in terms of the free-fall time of the initial cloud (top) or years (bottom).}
\label{BBB2003sfrate}
\end{figure}

In the lefthand panels of Figure \ref{global}, we present snapshots of the global evolution of the BBB2003 RT0.5 calculation.  The left most panels (using the red-yellow-white colour scale) display the column density, while the centre-left panels (using the blue-red-yellow-white colour scale) display the mass weight temperature in the cloud.  Figure \ref{BBB2003core} shows the evolution of the main dense core from the BBB2003 RT0.5 calculation in much greater detail than Figure \ref{global}.  These snapshots are shown at the same times as the equivalent figures in the original BBB2003 paper.  An animation comparing the original calculation with the radiation hydrodynamical calculation can be downloaded from http://www.astro.ex.ac.uk/people/mbate/~. 

Comparison of the snapshots and/or the animations shows that the barotropic and radiation hydrodynamical calculations diverge quickly on small scales.  In the original calculation, the first protostar to form is surrounded by a massive circumstellar disc that quickly fragments into three more objects -- two brown dwarfs and a low-mass star.  With radiation hydrodynamics, this massive disc does not fragment.  The accretion luminosity released as gas falls onto the disc and then spirals in towards the central protostar is sufficient to heat the disc and prevent its fragmentation.  This is one of the two main differences between the original calculation using the barotropic equation of state and the radiation hydrodynamical calculation -- discs and dense filaments close to existing protostars are inhibited from fragmenting by the radiative feedback due to the accretion luminosity released by the low-mass stars and brown dwarfs.  This is not surprising.  \citet{WhiBat2006} previously showed that replacing the barotropic equation of state with radiative transfer can lead to temperatures up to an order of magnitude higher near young low-mass protostars and, thus, potentially strongly inhibits fragmentation, while \citet{Krumholz2006} made a similar argument analytically.  \citet{Rafikov2005}, \citet{MatLev2005}, \citet{KraMat2006}, and \citet{WhiSta2006} have all pointed out that accurate treatments of radiative transfer are likely to significantly decrease disc fragmentation.  Finally, we note that the temperature field surrounding the newly-formed protostars varies significantly on timescales of hundreds to thousands of years.  In the animations, this can be seen as `flicking' of the temperature field.  These temperature variations are due to variations in the accretion rates of the protostars and their discs, particularly when protostars undergo dynamical interactions that perturb their discs.

The radiative feedback primarily affects fragmentation on length scales $\lsim 500$ AU.  As can be seen by comparing the snapshots in Figure \ref{BBB2003core} with the equivalent figures in BBB2003, the larger scales are unaffected, at least initially.  In Figure \ref{sepform}, we plot the cumulative distribution of the distances between each object and its closest other protostar at the time of formation of the object (i.e., the time a sink particle is inserted).  The thin solid line gives the distribution for the original barotropic BBB2003 calculation, while the thick solid line gives the distribution for the BBB2003 RT0.5 calculation.  Comparison of the two distributions shows that the fraction of objects forming within 100 AU of an existing protostar is twice as large without radiative feedback.  Similarly, more than 80\% of the objects form within 1000 AU of an existing protostar in the barotropic calculation while this percentage is reduced to 50\% with radiative feedback.  The small-scale effect of radiative feedback is in contrast to the effects of magnetic fields \citep[see][who performed simulations similar to BBB2003 with a range of magnetic field strengths]{PriBat2008} which affect the cloud's structure on both large and small scales.  Eventually, however, radiative feedback begins to alter structures on large scales {\em indirectly} because the chaotic dynamical interactions between protostars and ejections of stars and brown dwarfs from the cloud produce different gravitational potentials and move the gas distributions in different ways.

BBB2003 also presented snapshots of the evolution in two lower-mass dense cores that produced 7 and 5 objects, respectively.  In each core, all but one of these objects formed via the fragmentation of a disc surrounding an protostar.  With radiation hydrodynamics, these two cores produce 4 and 1 objects, respectively.  These two dense cores are shown at the end of the calculation in Figures \ref{BBB2003core23} for comparison with BBB2003.

Although radiative feedback strongly reduces the fragmentation on small length scales, this does not prohibit the formation of binary and multiple systems.  With only 13 objects produced by the calculation (Table \ref{table1}), it is impossible to discuss multiplicity statistics.  However, we note that when the calculation is stopped there is one system of six objects (two 2-AU binaries orbiting each other at 20 AU and this system is orbited by a 27-AU binary at 430 AU), one quintuple system (a 17-AU binary with companions orbiting at 65, 234, and 13000 AU), and two single stars.  As was found by \cite{BatBonBro2002b}, although two objects can form well separated from each other, the combination of dynamical interactions, gas accretion, and interactions with circumbinary and circum-multiple discs is very effective at producing multiple systems and even close binaries.  For example, the second and third objects to form in the calculation are initially separated by more than 1600 AU and yet at the end of the calculation they comprise one of the two 2-AU binaries.

In Figure \ref{BBB2003sfrate}, we plot the final mass of each object versus the time of its formation (i.e. the time that a sink particle was inserted) for the original BBB2003 calculation (top panel), the BBB2003 RT5 calculation (centre panel) and the BBB2003 RT0.5 calculation (lower panel).  The second main difference between the barotropic calculation and the radiation hydrodynamical calculations is that after the first burst of star formation in the main dense core ($t=1.03-1.13~t_{\rm ff}$ for each of the simulations, which is approximately the dynamical timescale of the dense core) there is a second burst of star formation in the barotropic calculation ($t=1.25-1.35~t_{\rm ff}$) but not in the radiation hydrodynamical simulations (BBB2003 RT5 has a weak burst at $t \approx 1.27~t_{\rm ff}$, but only one further object is formed in the main dense core of BBB2003 RT0.5).  Radiative feedback is responsible for almost shutting off the production of new stars in the main dense core.  Whereas stellar feedback is often considered one of the main agents in terminating the star formation process, the feedback invoked is usually that from massive stars, not low-mass stars (e.g. the formation of HII regions, strong winds, and supernova explosions).  The calculations presented here show that even the radiative feedback from low-mass star formation is enough to almost terminate the production of new stars in dense molecular cores with masses $\lsim 10$ M$_\odot$.  It does not destroy the dense molecular core or stop the existing protostars from accreting more gas, but it essentially turns off the production of new objects by raising the temperature of the bulk of the dense core so that few new gravitationally-unstable condensations are formed.

\begin{figure*}
\centering \vspace{-0.0cm}
    \includegraphics[width=15.8cm]{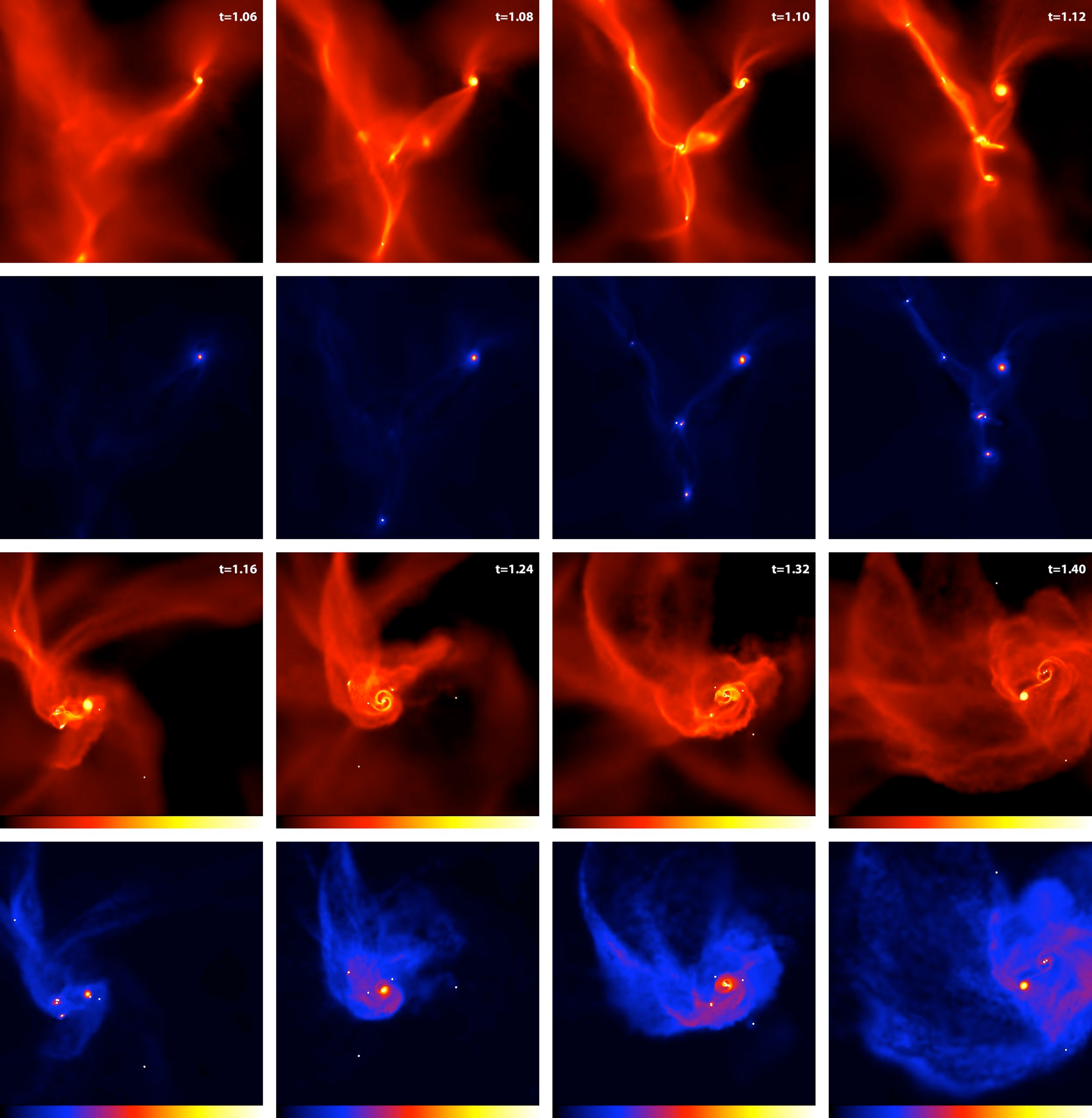}
\caption{The star formation in the main dense core of the BBB2003 RT5 calculation. These images are identical to those from the BBB2003 RT0.5 calculation in Figure \ref{BBB2003core} except that the calculation here uses sink particle accretion radii that are an order of magnitude larger (5 AU rather than 0.5 AU).  The early evolution of the cloud (top panels) is relatively independent of the sink particle parameters, but it can be seen that the gas near the prototars is slightly cooler with the larger accretion radii because a larger fraction of the accretion luminosity is neglected.  Later (lower panels), the two calculations slowly diverge.  There is slightly more fragmentation in the larger accretion radius calculation (but still substantially less than in the original barotropic BBB2003 calculation). }
\label{BBB2003_RT5core}
\end{figure*}

\subsubsection{Sink particles with $r_{\rm acc}=5$ AU}

\begin{figure*}
\centering
    \includegraphics[width=6cm]{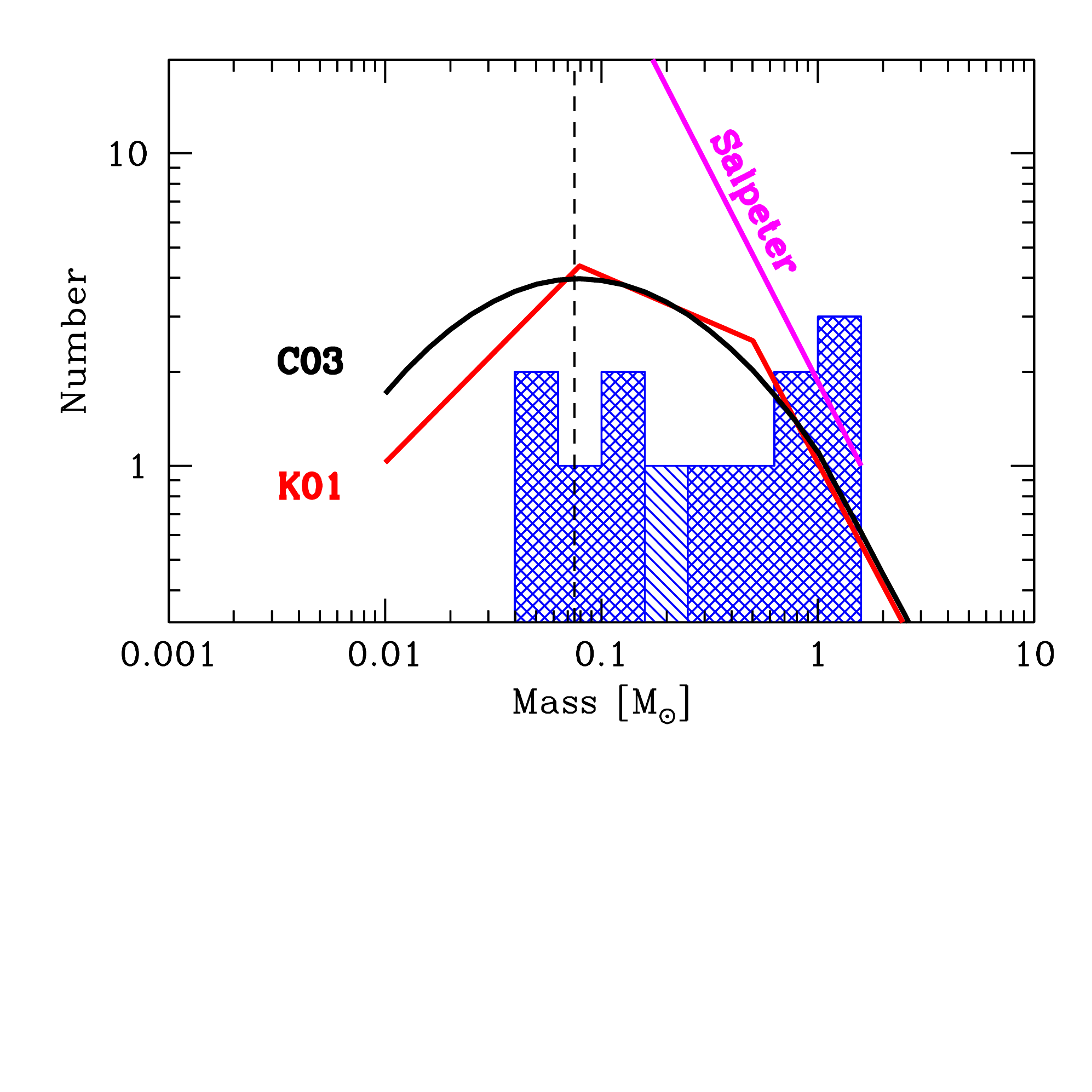} \hspace{-0.3cm}
    \includegraphics[width=6cm]{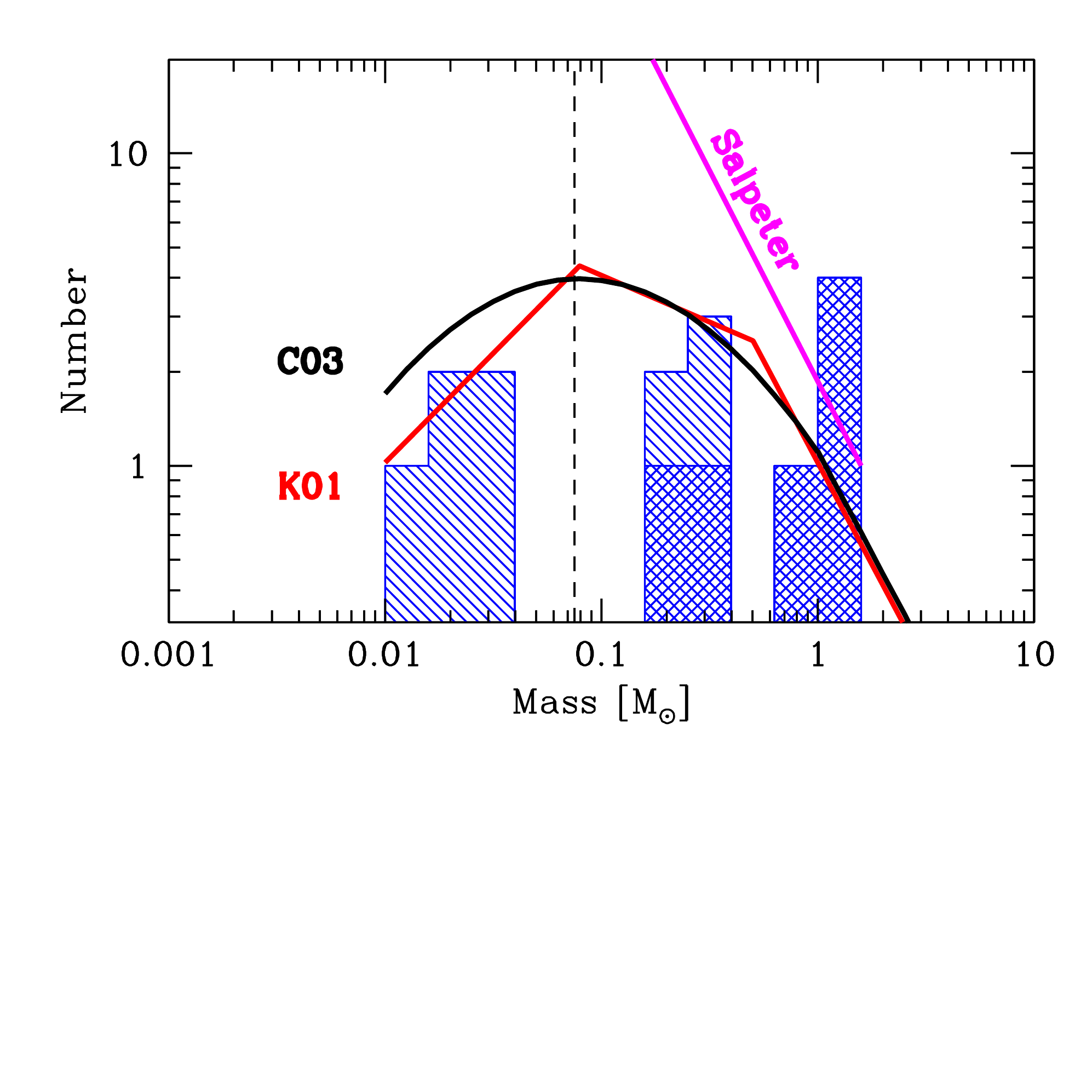} \hspace{-0.3cm}
    \includegraphics[width=6cm]{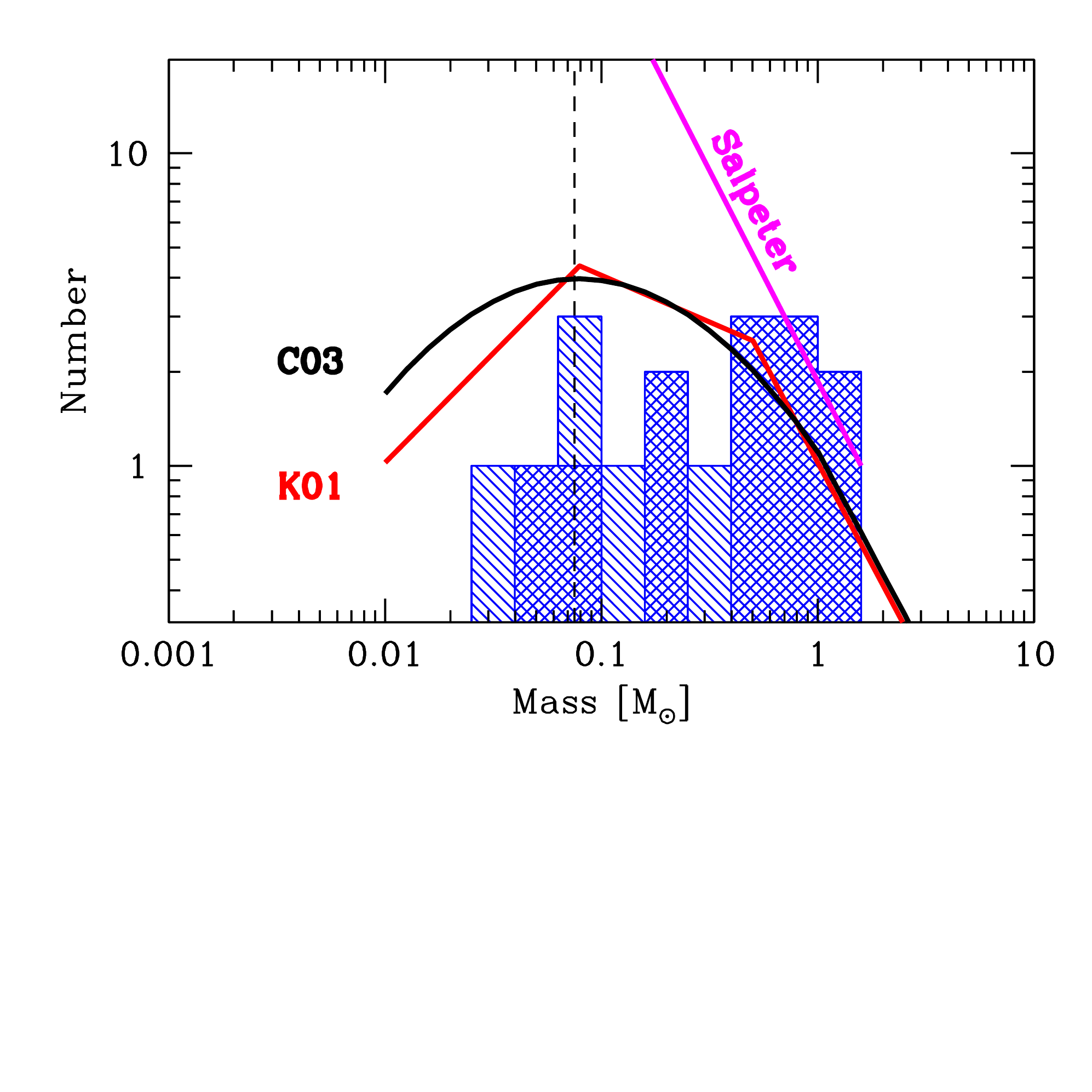} \vspace{-2.3cm}
\caption{Histograms giving the differential initial mass functions of the 13, 15, and 17 stars and brown dwarfs that had been produced by the end of the BBB2003 RT0.5, BBB2003 RT5, and BB2005 RT0.5 calculations, respectively.  The single hashed region gives objects that have stopped accreting while the double hashed region gives those objects that are still accreting.  Parameterisations of the observed IMF by \citet{Salpeter1955}, \citet{Kroupa2001} and \citet{Chabrier2003} are given by the magenta line, red broken power law, and black curve, respectively.}
\label{imfdiff}
\end{figure*}

In Figure \ref{BBB2003_RT5core}, we present snapshots of the evolution of the main dense core from the BBB2003 RT5 calculation (i.e. the same as above, but with sink particle accretion radii 10 times larger).  These snapshots should be compared with those in Figure \ref{BBB2003core}.  The initial evolution is very similar to that obtained with $r_{\rm acc}=0.5$~AU but it can be seen that the gas temperatures are slightly lower due to the fact that less accretion luminosity is injected into the calculation.  From the snapshots and Figure \ref{BBB2003sfrate} it can be seen that up until $t \approx 1.12~t_{\rm ff}$ the two calculations evolve in a similar manner.  However, the 5th object to form in BBB2003 RT5 is prevented from forming in the BBB2003 RT0.5 calculation due to the slightly higher gas temperatures in the central region visible in the $t=1.08$ and $1.12~t_{\rm ff}$ panels of Figures \ref{BBB2003core} and \ref{BBB2003_RT5core}.  In BBB2003 RT5, this object is formed only 24 AU from an existing protostar.  Subsequently in this central region, only one of the 8th, 9th, and 10th objects to form in BBB2003 RT5 (all within 350 AU of existing protostars) manages to collapse in the BBB2003 RT0.5 calculation.  Thus, it can be seen that neglecting the accretion luminosity emitted from within 5 AU of the protostars does increase the amount of fragmentation (although even this limited radiative feedback still reduces the number of objects formed by more than a factor of 3 compared to the barotropic equation of state).  Whether or not the 0.5 AU accretion radii calculation is converged in the sense that using smaller accretion radii would not change the number of fragments is, of course, hard to tell.  However, the small accretion radii calculation only forms two fewer objects than the large accretion radius calculation with the enhanced radiative feedback (Table \ref{table1}), and only two protostars form closer than 100 AU from existing protostars in the small accretion radii calculation (at 23 and 30 AU separations).  Thus, modelling the accretion flows right down to the surfaces of the protostars is unlikely to decrease the number of objects much further.

Beyond $t=1.14~t_{\rm ff}$, the calculations diverge because of the different numbers of objects and their chaotic dynamical interactions and ejections.  Interestingly, this even affects the star formation in the two lower-mass dense cores $\sim 2\times 10^4$ AU from the main dense core.  In the smaller accretion radii calculations each of these two dense cores begin forming their protostars $\approx 10,000$ years earlier than in the larger accretion radii calculations.  This is presumably due to slight perturbations of the lower-mass dense cores caused by the differing gravitational potentials due to the different evolutions of the main dense core.

\subsubsection{The initial mass functions}

\begin{figure}
\centering
    \includegraphics[width=8.4cm]{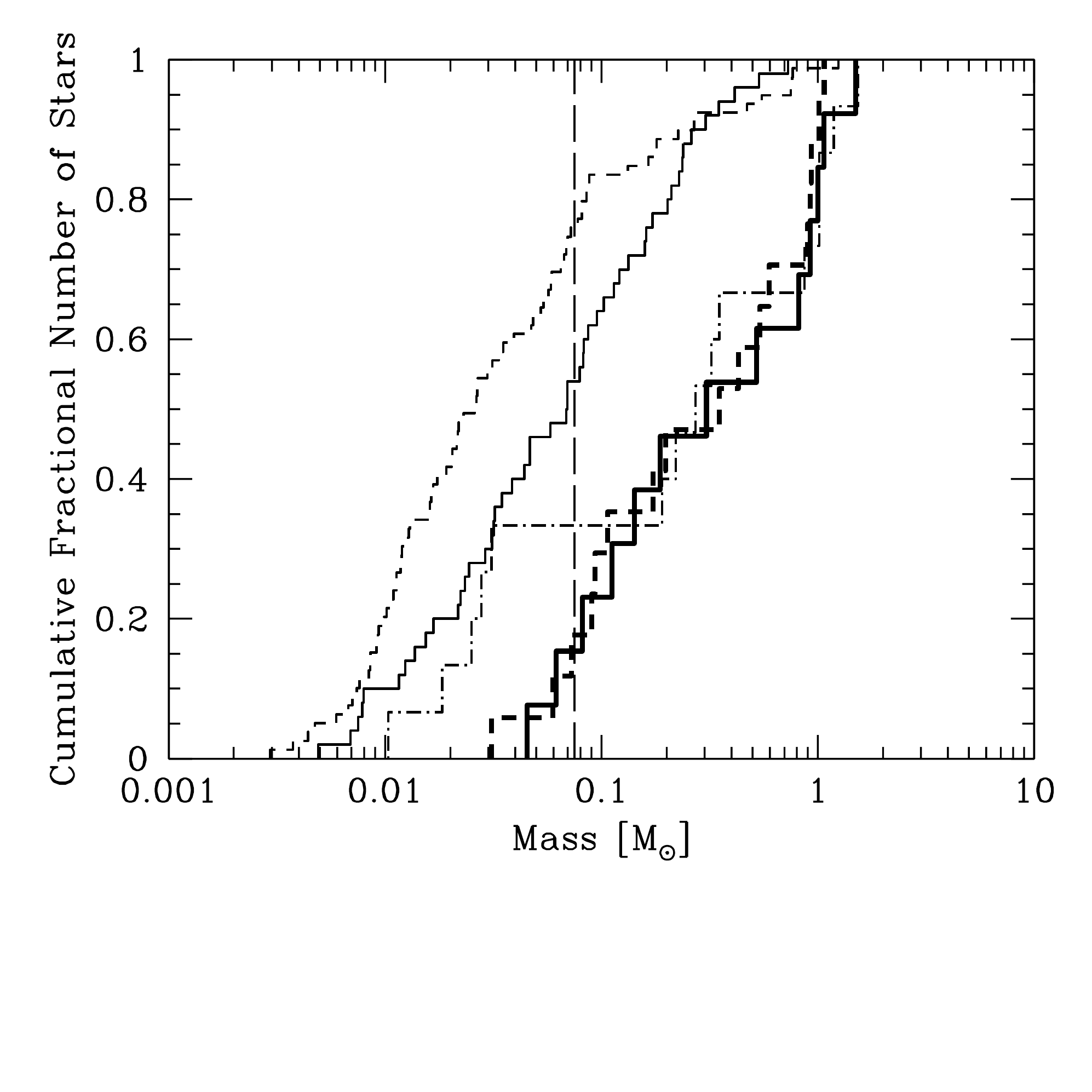}\vspace{-1.5cm}
\caption{The cumulative IMFs produced by all 5 calculations discussed in this paper.  The previously published IMFs from BBB2003 and BB2005 using a barotropic equation of state are given by the thin solid line and thin dashed line, respectively.  The radiation hydrodynamical calculations presented here are BBB2003 RT0.5 (thick solid line), BB2005 RT0.5 (thick dashed line) and the large accretion radius BBB2003 RT5 calculation (thin dot-dashed line).  The vertical long-dashed line denotes the boundary between brown dwarfs and stars.  It is clear that the radiation hydrodynamical calculations produce IMFs with a larger characteristic mass and far fewer brown dwarfs and low-mass stars than the original barotropic calculations.  Furthermore, whereas BBB2003 and BB2005 showed a clear dependence of the characteristic stellar mass on the initial Jeans mass of the molecular clouds (BB2005 began with a denser cloud with a Jeans mass 3 times lower that produced a median stellar mass 3.04 times lower than BBB2003), when radiative feedback is included there is no significant dependence of the IMF on cloud density and the initial Jeans mass.  A Kolmogorov-Smirnov (K-S) test on the BBB2003 RT0.5 and BB2005 RT0.5 distributions gives a 99.97\% probability that the two IMFs were drawn from the same underlying distribution (i.e. they are statistically indistinguishable).  By comparison, the two IMFs from the original barotropic calculations had only a 1.9\% probability of being drawn from the same underlying distribution.}
\label{imfcum}
\end{figure}

The effect of radiative feedback in terminating the production of new objects within the dense cores after one dynamical timescale and inhibiting the fragmentation of discs and filaments near existing protostars has a tremendous effect on the number of objects formed and the final distribution of stellar masses.  Table \ref{table1} summaries the numbers of stars and brown dwarfs formed, their combined mass, and their mean and median masses. The original barotropic BBB2003 calculation produced 50 stars and brown dwarfs in $1.40~t_{\rm ff}$.  However, BBB2003 RT0.5 only produced 13 objects in the same time and even BBB2003 RT5 with less accretion luminosity from the protostars produced only 15 objects.  Thus, the inclusion of radiative feedback has cut the number of objects produced by a factor of $\approx 4$.  In addition, whereas the original calculation produced a similar number of stars and brown dwarfs the ratio of brown dwarfs to stars is 1:3 for BBB2003 RT5.  For BBB2003 RT0.5, the ratio is less than 1:5 and both objects with brown dwarf masses are still accreting when the calculation is stopped.  

The much lower fraction of brown dwarfs is due to both the inhibiting of the fragmentation of discs and nearby filaments (because such objects are frequently ejected through dynamical interactions, terminating their accretion before they have been able to accrete much mass) and the suppression of new objects formed in the dense cores after a dynamical time.  In the latter case, it can be seen in Figure \ref{BBB2003sfrate} that there is a higher fraction of brown dwarfs amongst objects formed later in the barotropic calculation than those formed earlier (top panel).  Objects that form later must compete with the higher-mass protostars for the available gas.  Usually they lose, either being dynamically ejected from the dense core or at least having their velocities increased relative to the gas so that their accretion rates drop (Bondi-Hoyle accretion is proportional to $1/v^3$).  Ejections producing brown dwarfs and low-mass stars still occur, but they are much less common with the inclusion of radiative feedback than they were in the barotropic calculation.

Although the number of objects is decreased by the inclusion of radiative feedback, the amount of gas that has been converted into stars at $t=1.40~t_{\rm ff}$ is actually about 15\% greater compared to the barotropic calculation (see Table \ref{table1}).

The overall result of all of these effects is that the characteristic mass of the IMF moves to higher masses with the inclusion of radiative feedback and fewer brown dwarfs and low-mass stars are produced.  Comparing BBB2003 RT0.5 with BBB2003, the mean and median masses have increased by a factor of $4.4$ to $\approx 0.5$ M$_\odot$ and $\approx 0.3$ M$_\odot$, respectively (Table \ref{table1}).

Unfortunately, with so few objects it is not really possible to plot meaningful differential IMFs, but for completeness we include these in Figure \ref{imfdiff} in a similar manner to the original BBB2003 paper.   The IMFs are compared with the parameterisations of the observed IMF given by \citet{Chabrier2003}, \citet{Kroupa2001}, and \citet{Salpeter1955}.

In Figure \ref{imfcum}, we plot the cumulative IMFs produced by the original barotropic calculations and the three radiation hydrodynamical simulations presented here.  The BBB2003-type calculations are given using the solid lines (thin for the barotropic equation of state and thick for BBB2003 RT0.5) and the dot-dashed line for the larger accretion radius BBB2003 RT5 calculation.  The shift of the IMF to higher masses with the inclusion of radiative feedback is clear.  A Kolmogorov-Smirnov (K-S) test comparing the original calculation with BBB2003 RT0.5 gives only a 1.7\% probability that the two IMFs are drawn from the same underlying population (i.e. they are different).  We note that a K-S test comparing BBB2003 RT5 with BBB2003 RT0.5 shows they are indistinguishable (there is a 35\% probability they are drawn from the same population).

\subsection{BB2005 initial conditions}

\begin{figure*}
\centering \vspace{-0.0cm}
    \includegraphics[width=17.7cm]{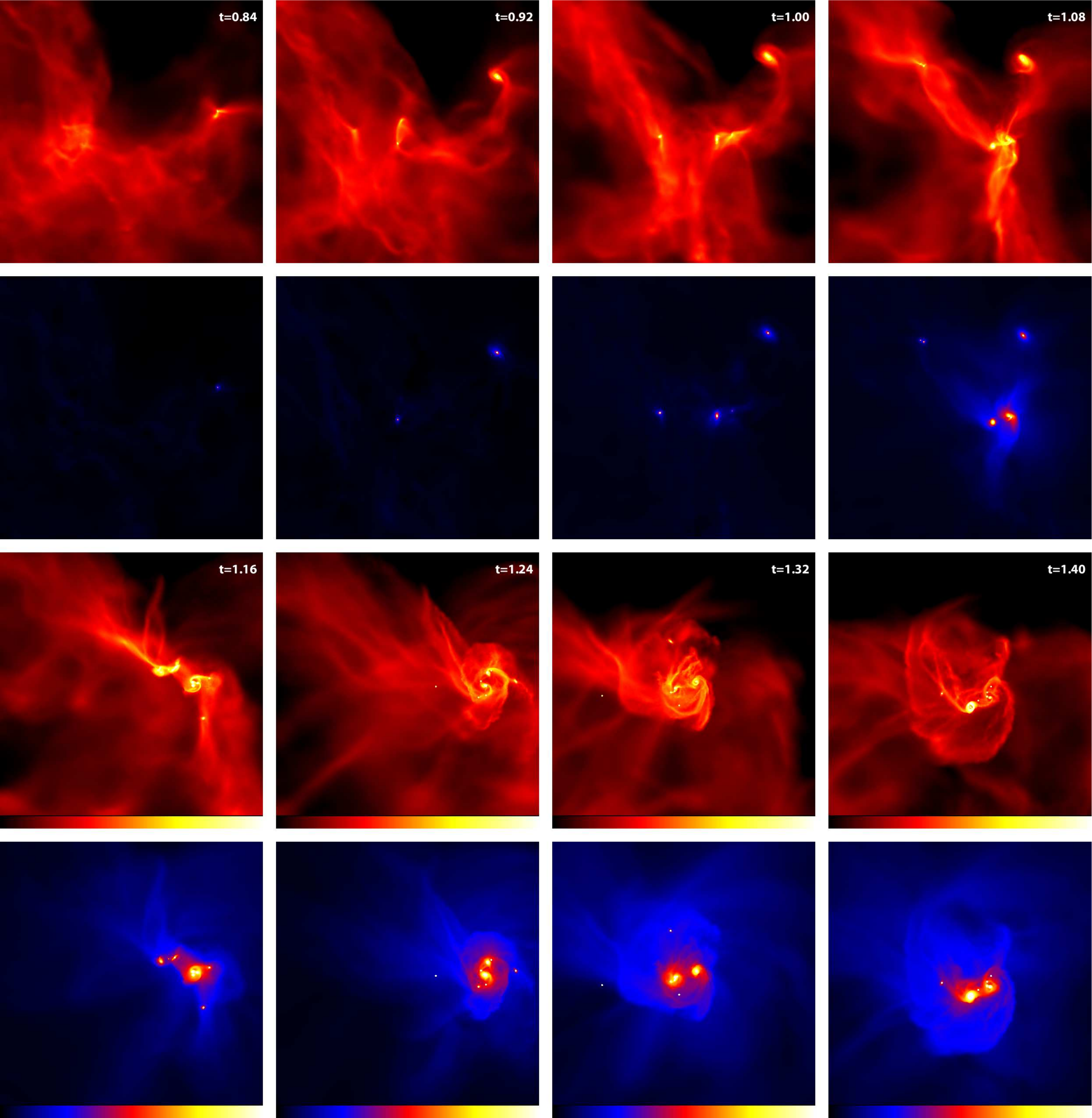}
\caption{The star formation in the main dense core of the BB2005 RT0.5 calculation. The first object forms at $t=0.850~t_{\rm ff}$. Large gaseous filaments collapse to form single objects and multiple systems. These objects fall together to form a small group. Dynamical interactions within the group eject a few objects. Radiative feedback from the accreting protostars heats the gas in the dense core.  Each panel is 0.025 pc (5150 au) across. Time is given in units of the initial free-fall time of $6.34\times 10^{4}$ yr. The red-yellow-white panels show the logarithm of column density, N, through the cloud, with the scale covering $-0.2 < \log N < 2.5$ with N measured in g~cm$^{-2}$.  The blue-red-yellow-white panels show the logarithm of mass weighted temperature, $T$, through the cloud with the scale covering $9-300$~K.}
\label{BB2005core}
\end{figure*}

We turn now to the second set of initial conditions, those from BB2005 which are identical to those of BBB2003 except that the initial cloud has 9 times the density due to a smaller radius, and a higher Mach number so that for both types of initial conditions the kinetic and gravitational potential energies have the same magnitude (see Table \ref{table1}).  Since the cloud has the same initial temperature, the mean thermal Jeans mass is a factor of 3 lower ($1/3$~M$_\odot$ rather than 1~M$_\odot$).  In the original barotropic calculations, the lower initial Jeans mass resulted in a proportional lowering of the median stellar mass of the objects produced, showing that the characteristic stellar mass is set by the initial Jeans mass in such simulations.

The radiation hydrodynamical version of the BB2005 calculation (BB2005 RT0.5) uses sink particles with the more accurate small accretion radii ($r_{\rm acc}=0.5$ AU).  Again, we do not begin the radiation hydrodynamical simulation from the initial conditions because the early evolution of the cloud is essentially isothermal.  We begin calculation BB2005 RT0.5 from the last dump of BB2005 before the density exceeded $9 \times 10^{-16}$~g~cm$^{-3}$ at $t=0.692~t_{\rm ff}$.  
In the original BB2005 calculation the first sink particle was inserted at $t=0.824~t_{\rm ff}$, some $8.4\times 10^3$ years later.  Using radiation hydrodynamics again delays the insertion of the first sink particle (for the same reason as given in Section \ref{BBB2003text}) until $t=0.850~t_{\rm ff}$.

In the righthand panels of Figure \ref{global}, we present snapshots of the global evolution of the BB2005 RT0.5 calculation.  Figure \ref{BB2005core} shows the evolution of the main dense core from the BB2005 RT0.5 calculation in much greater detail than Figure \ref{global}.  Again, these are shown at the same times as the equivalent figures in the original BB2005 paper to allow direct comparison.  Again, an animation comparing the original calculation with the radiation hydrodynamical calculation can be downloaded from http://www.astro.ex.ac.uk/people/mbate/~.

As with the BBB2003-type initial conditions, the barotropic and radiation hydrodynamical calculations diverge quickly on small scales.  In the barotropic calculation, the first two objects form from two dense clumps separated by approximately 100 AU, one object forming 700 years after the other.  This binary accretes a massive circumbinary disc which later fragments to form a third object.  In the radiation hydrodynamical simulation, the accretion luminosity from the first object inhibits the formation of the second object and the massive disc which subsequently forms around this single protostar is too hot to fragment.  Thus, whereas three objects were formed in this small region of the original simulation, the radiation hydrodynamical simulation forms only a single protostar.  This pattern continues throughout the early evolution -- fragmentation on length scales $\gsim 300$ AU is unaffected by the accretion luminosity emitted by existing protostars, but most of the fragmentation on smaller scales (of either nearby dense filaments or massive circumstellar discs) is stopped.  Figure \ref{sepform}, shows that more than 3/4 of the objects form within 300 AU of an existing protostar in the original barotropic BB2005 calculation (thin dashed line) whereas this fraction is reduced to less than 1/3 with radiative feedback (thick dashed line).  As with the BBB2003-type initial conditions, the radiative feedback eventually begins to alter structures on larger scales indirectly because of the differing numbers of protostars and their dynamical interactions ($t \gsim 1.20~t_{\rm ff}$).  

As discussed above for the first type of initial conditions, the reduction of small-scale fragmentation due to the radiative heating does not stop the production of binary and multiple objects.  At the end of the calculation there are 2 single very-low-mass objects (masses $<0.1$~M$_\odot$), a triple system consisting of a 150-AU binary with a wide (8400-AU) companion, and a small bound cluster of 12 objects, including three binaries with semi-major axes of less than 3 AU, one of which has a close third companion with a semi-major axis of 20 AU.

\begin{figure}
\centering
    \includegraphics[width=7.5cm]{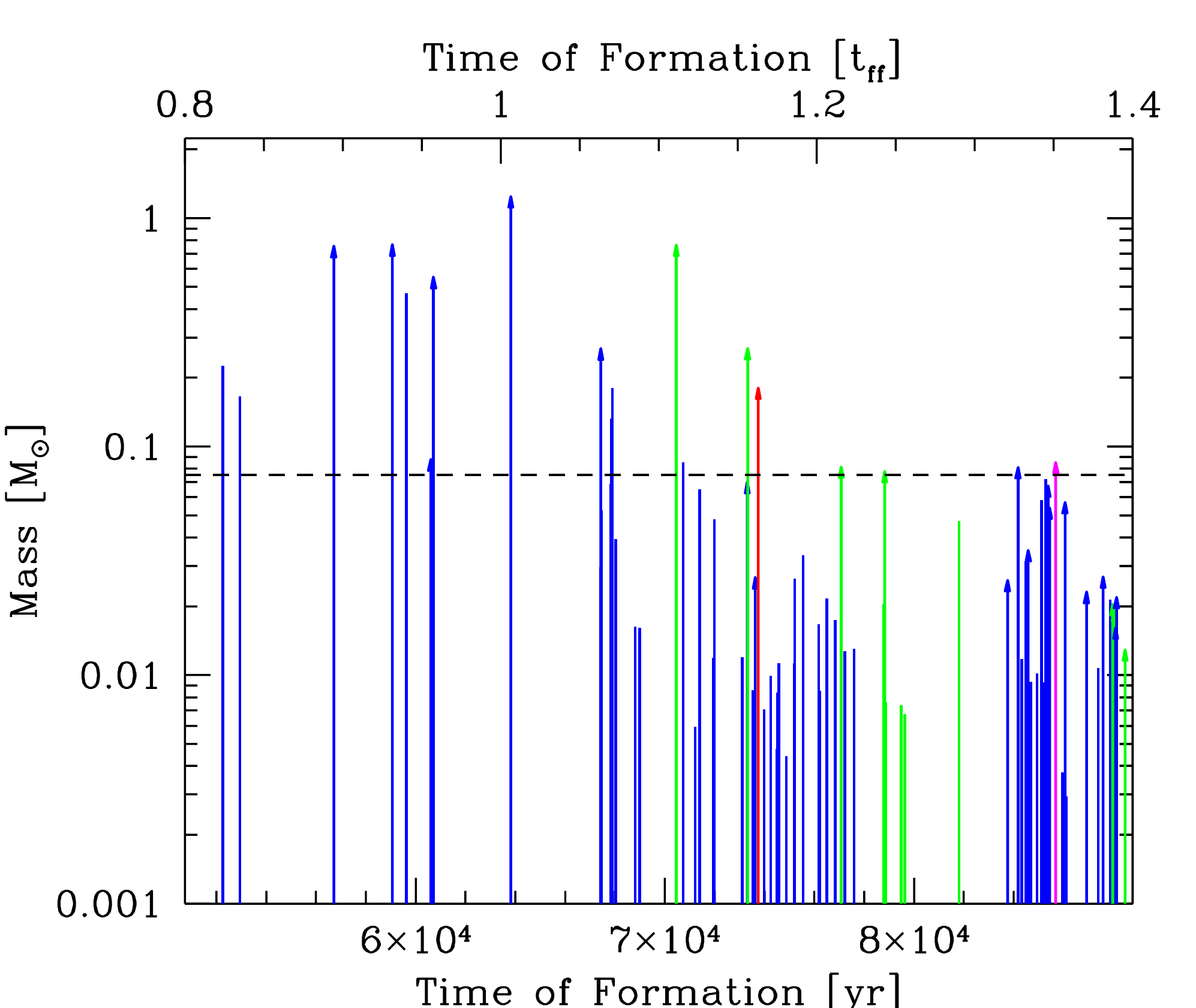}\vspace{0.0cm}
    \includegraphics[width=7.5cm]{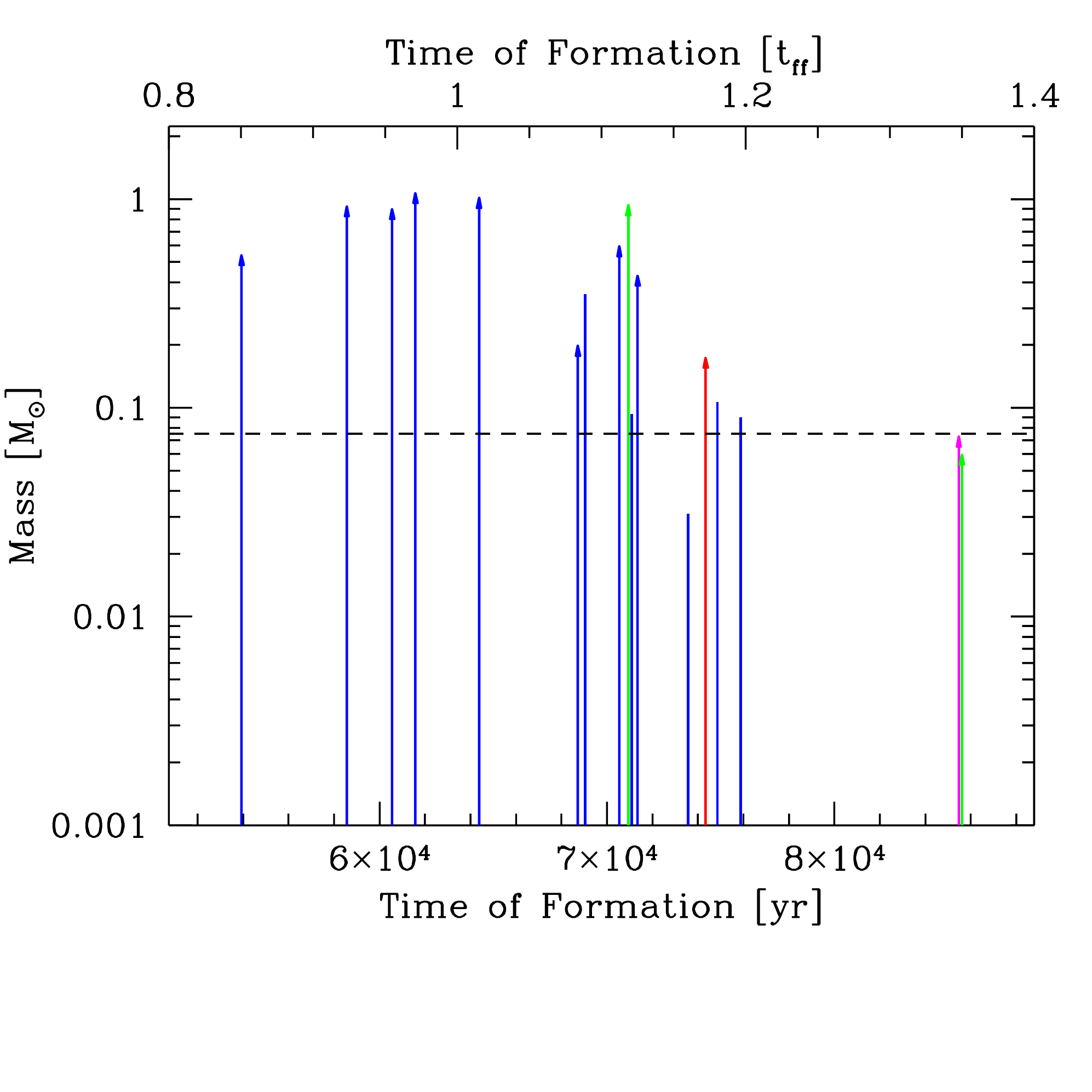}\vspace{-1.0cm}
\caption{Time of formation and mass of each star and brown dwarf at the end of original BB2005 calculation (top) and the radiation hydrodynamical calculation BB2005 RT0.5 (bottom).  Again, radiative feedback does not alter the initial phase of star formation too much, but most of the continued star formation that occurred in the original calculation is inhibited by the radiative feedback.  Objects that form in the main dense core are denoted by blue lines.  Objects that form in the lower-mass second, third, and fourth dense cores are denoted by green, red, and magenta lines, respectively.  Objects that are still accreting significantly at the end of the calculation are represented with vertical arrows.  The horizontal dashed line marks the star/brown dwarf boundary.  Time is measured in terms of the free-fall time of the initial cloud (top) or years (bottom).}
\label{BB2005sfrate}
\end{figure}

In Figure \ref{BB2005sfrate}, we plot the final mass of each object versus the time of its formation (i.e. the time that a sink particle was inserted) for the original BB2005 calculation (top panel) and the BB2005 RT0.5 calculation (lower panel).  As with the first type of initial conditions, the radiative feedback terminates the production of new objects in the main dense core well before the simulation is stopped.  By contrast, in the barotropic calculation, production of stars and brown dwarfs in the main dense core continues until the calculation is stopped, with only a brief pause at $t=1.22-1.30~t_{\rm ff}$.  With radiative feedback the main dense core ceases production of new objects at $t=1.20~t_{\rm ff}$ because the gas in the main dense core has been heated by the embedded protostars.

In the original BB2005 calculation, star formation proceeded in three lower-mass dense cores in addition to the main dense core.  This is also true of the radiation hydrodynamical simulation.  Two of these dense cores (with masses $0.2-0.3$~M$_\odot$ when the sink particles are inserted) produce single protostars in both the barotropic and radiation hydrodynamical calculations.  The remaining dense core (with a mass of $\approx 1$~M$_\odot$ when star formation begins) formed 12 objects in the barotropic calculation due to a combination of disc fragmentation and filament fragmentation.  With radiation hydrodynamics, this core only produces 2 objects due to fragmentation of a filament.

\subsubsection{The initial mass function}

As with the lower-density cloud, the radiative feedback dramatically decreases the number of objects formed (see Table \ref{table1}).  The original BB2005 calculation produced 79 stars and brown dwarfs in 1.40~$t_{\rm ff}$, whereas with radiation hydrodynamics only 17 objects form.  Whereas the barotropic calculation produced three times as many brown dwarfs as stars, this ratio is more than inverted when radiative feedback is included, with more than 4 times as many stars as brown dwarfs (including one brown dwarf that has had its accretion terminated).  Again, this lower proportion of brown dwarfs is due to the decrease in disc fragmentation and the collapse of nearby filaments, as well as the cessation of fragmentation in the main dense core at $t=1.20 t_{\rm ff}$ due to the heating of the core by the embedded protostars (Figure \ref{BB2005sfrate}).

The amount of gas converted into stars and brown dwarfs at the end of the calculations is similar in the barotropic and radiation hydrodynamical calculations (only differing by 4\%: Table \ref{table1}), but as with the BBB2003-type initial conditions the large decrease in the number of objects produced moves the characteristic mass of the IMF to much higher masses.  The mean and median masses are increased by factors of 4.5 and 15, to $\approx 0.5$~M$_\odot$ and $\approx 0.3$~M$_\odot$, respectively.

The differential and cumulative IMFs are displayed in Figures \ref{imfdiff} (right panel) and \ref{imfcum} (thick dashed line), respectively, in a similar manner to those given in the original BB2005 paper.  The barotropic cumulative IMF is also plotted in Figure \ref{imfcum} (thin dashed line), clearly displaying the increase in the typical stellar mass.  A K-S test comparing the original barotropic calculation with BB2005 RT0.5 gives only an 0.0004\% (i.e., 1/250,000) probability that the two IMFs are drawn from the same underlying population (i.e. they are distinctly different).

\section{Discussion}
\label{discussion}

\subsection{The ratio of stars to brown dwarfs}

The observed initial mass function within the local region of our Galaxy is now reasonably well constrained, at least down to $\approx 0.03$ M$_\odot$.  Surprisingly, there is little evidence for variation of the IMF amongst star-forming regions, open clusters, and even globular clusters, leading to the idea that the IMF may be universal (at least for metallicities greater than $\sim 1/1000$ solar). Two frequently used parameterisations of the observed IMF are given by \citet{Kroupa2001} and \citet{Chabrier2003} (see Figure \ref{imfdiff}).  In particular, it is now generally accepted that the number of stars is larger than the number of brown dwarfs \citep{Chabrier2003, Greissletal2007, Luhman2007, Andersenetal2008}.  \citet{Andersenetal2008} find that the ratio of stars with masses $0.08-1.0$ M$_\odot$ to brown dwarfs with masses $0.03-0.08$ M$_\odot$ is $N(0.08-1.0)/N(0.03-0.08)\approx 5\pm 2$.

Previous hydrodynamical simulations using a barotropic equation of state have consistently produced more brown dwarfs than stars (BBB2003, BB2005, \citealt{Bate2005, Bate2009a}).  In the earlier calculations, this might have been explained away by small number statistics (the earlier calculations each produced fewer than 100 objects).  However, \cite{Bate2009a} recently performed a barotropic calculation of the collapse and fragmentation of a 500 M$_\odot$ cloud that produced 1254 stars and brown dwarfs.  While some of the stellar properties (e.g. the multiplicity as a function of primary mass) were in good agreement with observations, this calculation produced at least 50\% more brown dwarfs than stars.  The large number of objects produced leaves no doubt as to the statistical significance of this result.

For this reason, the decrease in the fraction of brown dwarfs to stars provided by the inclusion of radiative feedback is welcome.  Averaging over the two small accretion radii calculations presented here, the ratio of stars to brown dwarfs is found to be $\gsim\;$5.  Although the statistics are poor, this is at least in reasonable agreement with observations.  Obviously the next step is to perform simulations of more massive molecular clouds in order to improve the statistics as was done by \citet{Bate2009a} for the barotropic calculations.

Magnetic fields may also affect the mass function.  Recently, \citet{PriBat2007} showed that stronger magnetic fields generally inhibit disc formation and binary formation \citep[see also][]{HenFro2008,HenTey2008}.  \citet{PriBat2008} ran star cluster formation simulations similar to BBB2003, but with magnetic fields.  They found that the extra pressure support provided by magnetic fields generally decreased the rate of star formation and the importance of dynamical interactions between objects.  This leads to stronger magnetic fields producing a decrease in the ratio of brown dwarfs to stars (though again the total numbers of objects formed in the calculations were small, ranging from 15 to 69). Finally, \citet*{OffKleMcK2008} report that simulations of driven turbulence produce fewer low-mass objects than simulations with decaying turbulence.

\subsection{Comparison with the Taurus-Auriga star-forming region}

The Taurus-Auriga star-forming region is a large but diffuse molecular cloud spanning 30 pc and containing $2\times 10^4$ M$_\odot$ of molecular gas \citep{Goldsmithetal2008}.  Although the entire complex is much larger and more massive than the simulations presented here, many of the recently formed stars in Taurus are contained within 6 or 7 groups each containing $\approx 10$ members and measuring $\approx 2$ pc across \citep{Gomezetal1993}.  These groups may have originated from dense cores similar to the two main dense cores modelled in the calculations presented here which form 8 and 13 objects.  Once dispersed, these groups would appear very similar to the Taurus groups.  The IMF of the Taurus cloud \citep{Luhmanetal2003a} is also similar to the IMFs presented in Figure \ref{imfdiff} in the sense that they both seem to contain an excess of $\sim 1$ M$_\odot$ stars when compared to parameterisations of a universal IMF.  

Although the Taurus IMF and the simulation IMFs are {\em consistent} with a universal IMF due to the small number statistics, it may also be that low-density distributed star-forming regions that form stars in small groups rather than large clusters do preferentially form stars with a masses $\approx 1$ M$_\odot$.  In the simulations presented here, the small groups form in dense molecular cores with masses of $\approx 10$~M$_\odot$ that fragment into $\approx 10$ objects.  With a few objects dynamically ejected as brown dwarfs or low-mass stars before they have accreted much of the available mass, the remaining objects can each accrete to $\approx 1$ M$_\odot$ before the gas reservoir is exhausted.  This leaves a stellar mass distribution that is biased in favour of solar-type stars, with a few lower mass members.  Such an explanation for the IMF in Taurus was first presented by \cite*{GooWhiWar2004c} based on hydrodynamical simulations of isolated dense cores.  If this model is correct, the simulations presented here may be best compared with the small Taurus stellar groups.

\subsection{The characteristic mass of the IMF}

The calculations of BBB2003, BB2005, and \cite{Bate2005} tested the dependence of the IMF obtained from barotropic hydrodynamical simulations on various changes to the initial conditions and metallicity of molecular clouds.  The conclusion from these papers was that the characteristic mass of the IMF is set by the initial Jeans mass of the molecular clouds.  Other similar hydrodynamical calculations, that did not resolve brown dwarfs or close binaries, have led to similar conclusions \citep{KleBurBat1998,KleBur2000, Jappsenetal2005, BonClaBat2006}.  Current thinking is that this initial Jeans mass may be set by the thermodynamics of molecular gas at the transition from atomic line cooling to dust cooling which sets a characteristic density and Jeans mass \citep{Larson1985, Larson2005}.  Again, this appears to be backed up by hydrodynamical simulations showing that the peak of the IMF scales linearly with a change in the density at which this transition occurs \citep{Jappsenetal2005, BonClaBat2006}.  Even other models of the origin of the IMF predict that the characteristic mass of the IMF should scale linearly with the typical Jeans mass.  For example, the theory of \citet{PadNor2002} which proposes that the IMF originates from the mass spectrum of dense cores formed by supersonic turbulence predicts that the characteristic mass of the IMF scales linearly with the Jeans mass of the molecular cloud but also depends on the Mach number of the turbulence.

Recently, \cite{ElmKleWil2008} proposed that the reason the IMF appears to be a universal function in the local Universe is because this characteristic thermodynamic Jeans mass in molecular clouds is relatively insensitive to initial conditions such as molecular cloud density, the local radiation field, etc.  However, for the dense cores where stellar groups form this theory implies that temperature at dust-gas coupling increases with density as $n^{1/2}$.  If anything, observations tend to show the temperature in dense molecular gas  decreases with increasing density.  This theory also means that the IMF is universal because of the thermodynamics of gas and dust, not due to the star formation process itself.  It would be more elegant if star formation somehow regulated itself to provide a near universal IMF in the local Universe.

Given the wide range of possible theories for the origin of the IMF (see Section \ref{introduction}), little attention has been paid to the role of radiative feedback.  Dynamical feedback processes such as winds and outflows have been considered \citep{Shuetal1988, Silk1995, AdaFat1996}, but the effect of radiative feedback on stellar masses is usually only considered for massive stars in the context of setting a maximum stellar mass.  One exception is star formation in massive accretion discs surrounding supermassive black holes.  \cite{Nayakshin2006} proposes that radiative feedback from low-mass protostars forming in a gravitationally-unstable disc surrounding a supermassive black hole may heat the disc enough to stop further fragmentation.  Trapped in the disc, the few protostars that did manage to form before the disc was stabilised by the protostellar heating would then accrete the remaining gas becoming very massive.  \citeauthor{Nayakshin2006} proposes that this may explain the apparent top-heavy mass function of the stars orbiting the Sgr A$^*$ in the centre of our Galaxy \citep{NaySun2005}.  Although this case is very different to local star formation, it is a case where radiative feedback, even from low-mass protostars, may play a substantial role in setting stellar masses.  In what follows, we show that radiative feedback may not just set the characteristic stellar mass in this exotic case --  it may also be responsible for setting the characteristic stellar mass in typical star formation.

\subsubsection{The effect of radiative feedback on the characteristic mass}

In the barotropic calculations of BBB2003 and BB2005, the density of the initial molecular cloud was 9 times higher in the latter calculation leading to an initial mean thermal Jeans mass three times lower in the denser cloud.  The mass distributions of the stars and brown dwarfs produced by these calculations had median masses that scaled almost exactly with this change in the initial Jeans mass -- the median mass was 3.04 times smaller for the stellar cluster produced by the denser cloud.  K-S tests performed on the two distributions showed that there was less than a 1.9\% chance that they were drawn from the same underlying mass function.

However, as seen above, when the calculations are repeated with radiative feedback the resulting IMFs have a significantly larger characteristic mass and, more importantly, they are indistinguishable from each other (Figure \ref{imfcum}) despite the different initial mean thermal Jeans masses of the clouds.  A K-S test comparing the IMFs of the BBB2003 RT0.5 and BB2005 RT0.5 calculations gives a 99.97\% probability they are drawn from the same population.  This implies that {\em radiative feedback during the star formation process substantially weakens the dependence of the IMF on the initial Jeans mass in the molecular cloud and may be responsible for producing a nearly universal IMF}.

The general idea of how radiative feedback may increase the characteristic mass of the IMF is relatively simple to understand.  The Jeans length scales with temperature and density in a molecular cloud as 
\begin{equation}
\label{stdjeanslength}
\lambda_{\rm J}\propto T^{1/2}\rho^{-1/2},
\end{equation}
so the Jeans mass scales as 
\begin{equation}
\label{stdjeansmass}
M_{\rm J}\propto \rho \lambda_{\rm J}^3 \propto T^{3/2}\rho^{-1/2}.
\end{equation}
Thus, for the two types of initial conditions here which have the same temperature but a density that differs by a factor of 9, the initial Jeans mass and Jeans length are 3 times smaller in the denser cloud.  When the gas collapses, the fragments (protostars) that form are roughly separated by the Jeans length so that in the denser cloud they are closer together and, therefore, each has a smaller reservoir from which to accrete.  Essentially, a molecular cloud of mass $M_{\rm c}$ is broken into $M_{\rm c}/M_{\rm J}$ objects that, on average, accrete the mass available in the reservoir contained within a Jeans length and, thus, their characteristic mass is $\approx M_{\rm J}$  (some objects accrete more and some less due to competitive accretion, but the characteristic mass is $\approx M_{\rm J}$).  

However, radiative feedback changes this situation.  When the first fragment forms, this protostar heats the gas around it.  Nearby gas, which would otherwise have collapsed soon after the first object to form additional objects, may now be prohibited from collapsing because it is hotter and that gas is instead accreted by the already existing protostar.  Thus, the distance between neighbouring protostars is increased because of the radiative feedback and those that do form each have a larger reservoir of gas to accrete.  In other words, the {\em effective} Jeans length and Jeans mass (the mass that will end up in that object rather than collapsing to form a neighbouring object) has increased.  This explains the general increase in the characteristic mass.

\subsubsection{Dependence of the characteristic mass on initial conditions}

Understanding how the characteristic mass of the IMF may depend on variations in the initial conditions (e.g. the cloud's mean density or temperature) when radiative feedback is included is somewhat more complicated.  

Let us begin by assuming that the mean density of the molecular cloud is low enough that the cloud is optically thin to infrared radiation.  At a typical temperature of $10$~K for solar metallicity molecular gas the mean dust absorption opacity is $\kappa \approx 0.01$ cm$^{2}$~g$^{-1}$ so that for molecular hydrogen densities of $10^6$ cm$^{-3}$ an optical depth of unity is reached after $\approx 10$ pc, much larger than the typical distance between protostars in a typical star-forming region.  Even taking a higher density of $10^7$ cm$^{-3}$ and a higher temperature of 30~K with a corresponding mean opacity of $\kappa = 0.1$ cm$^{2}$~g$^{-1}$ this length scale is 0.1 pc.  For simplicity, and because the radiation hydrodynamical calculations presented in this paper have been performed using grey radiative transfer, we also assume that the dust absorption opacity is independent of wavelength.  The more general case, in which the wavelength dependence of the dust absorption opacity scales as $q_{\rm abs} \propto \lambda^{-\beta}$, is relatively easy to derive but the extra complication distracts from the main point of the discussion below.  

Under these assumptions, the temperature, $T$, of the gas and dust at radius, $r$, from a (spherically symmetric) protostar of luminosity, $L_*$, is given by
\begin{equation}
\label{temperature}
L_* = 4 \pi \sigma r^2 T^4,
\end{equation}
where $\sigma$ is the Stephan-Boltzman constant.

Now, a convenient definition of the Jeans length, $\lambda_{\rm J}$, is that it is the radius at which the sound speed of the gas equals the escape velocity of the mass enclosed within this radius
\begin{equation}
\label{escape}
c_{\rm s}^2 = \frac{\cal{R}}{\mu} T =\frac{GM}{\lambda_{\rm J}},
\end{equation}
where $c_{\rm s}$ is the sound speed, $\cal{R}$ is the gas constant, $\mu$ is the mean molecular weight, and the Jeans mass is
\begin{equation}
\label{massradius}
M_{\rm J}= \frac{4 \pi}{3} \rho \lambda_{\rm J}^3.
\end{equation}
On scales smaller than $\lambda_{\rm J}$, the gas is supported by pressure against collapse, while on larger scales the gas is unstable to collapse.
For the usual definition of the Jeans length and mass, the sound speed is taken to be a constant and by combining equations \ref{escape} and \ref{massradius} we obtain equations \ref{stdjeanslength} and \ref{stdjeansmass} .  

However, for a gas cloud that is internally heated by an embedded protostar we can use equations \ref{temperature}, \ref{escape}, and \ref{massradius} to obtain
\begin{equation}
\label{generalradius}
\lambda_{\rm eff} = \rho^{-2/5} L_*^{1/10}  \left(\frac{3~ \cal{R}}{4\pi~ \mu G}\right)^{2/5} \left(4 \pi \sigma\right)^{-1/10},
\end{equation}
and
\begin{equation}
\label{jeansmass}
M_{\rm eff} = \rho^{-1/5} L_*^{3/10} ~ \frac{4\pi}{3} \left(\frac{3~ \cal{R}}{4\pi~ \mu G}\right)^{6/5} \left(4 \pi \sigma\right)^{-3/10},
\end{equation}
where we have used $\lambda_{\rm eff}$ and $M_{\rm eff}$ to differentiate the {\em effective} Jeans length and mass due to radiative feedback from the standard (isothermal) definitions, $\lambda_{\rm J}$ and $M_{\rm J}$.
Thus, the effective Jeans mass in a molecular cloud containing accreting protostars is significantly less dependent on the density of the cloud than the usual Jeans mass, $M_{\rm J}$ (which scales $\propto \rho^{-1/2}$).  This dependence of the effective Jeans mass on density is so weak that it is consistent with the results of the radiation hydrodynamical simulations presented here in which we find no significant variation of the characteristic mass of the IMF with density.  Simulations that formed a much greater number of objects would be needed to detect such a weak dependence.  We also note that the dependence on density in equation \ref{jeansmass} is even weaker than that given by the theory of \cite{ElmKleWil2008}, and it has the opposite sign (i.e. the characteristic mass decreases with increasing density whereas \citeauthor{ElmKleWil2008} find that it should increase with increasing density).  Finally, as mentioned above, we have neglected the wavelength dependence of the dust absorption opacity.  Including this alters the powers of density and luminosity that appear in equations \ref{generalradius} and \ref{jeansmass} slightly.  For example, the scaling of the effective Jeans mass with density becomes $\rho^{-1/(5-\beta)}$ so that for dust with $\beta=1$ the effective Jeans mass scales as  $\rho^{-1/4}$.  This is still substantially weaker than the usual scaling of Jeans mass with cloud density.

\begin{figure}
\centering\vspace{-0.2cm}
    \includegraphics[width=6cm]{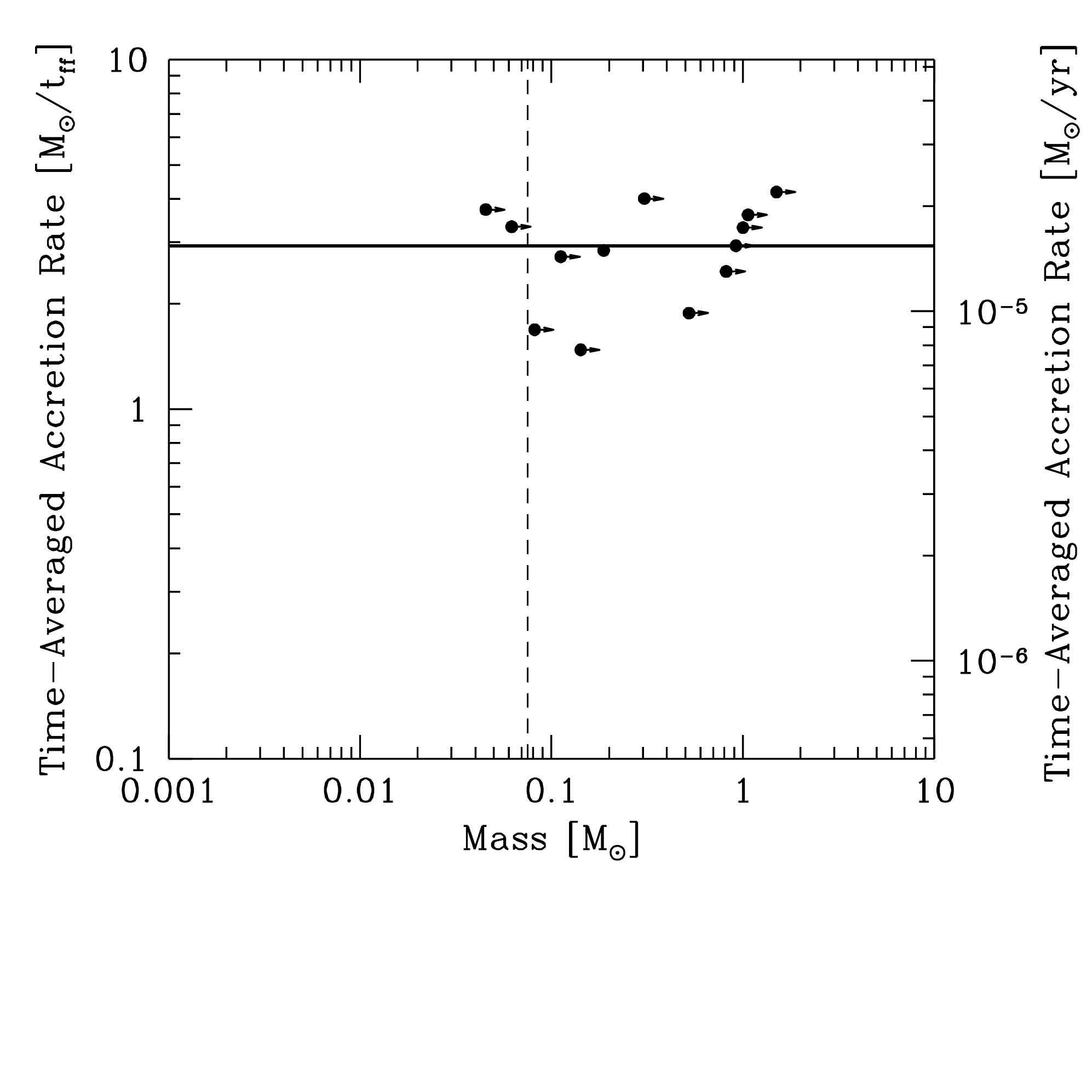}\vspace{-1cm}
    \includegraphics[width=6cm]{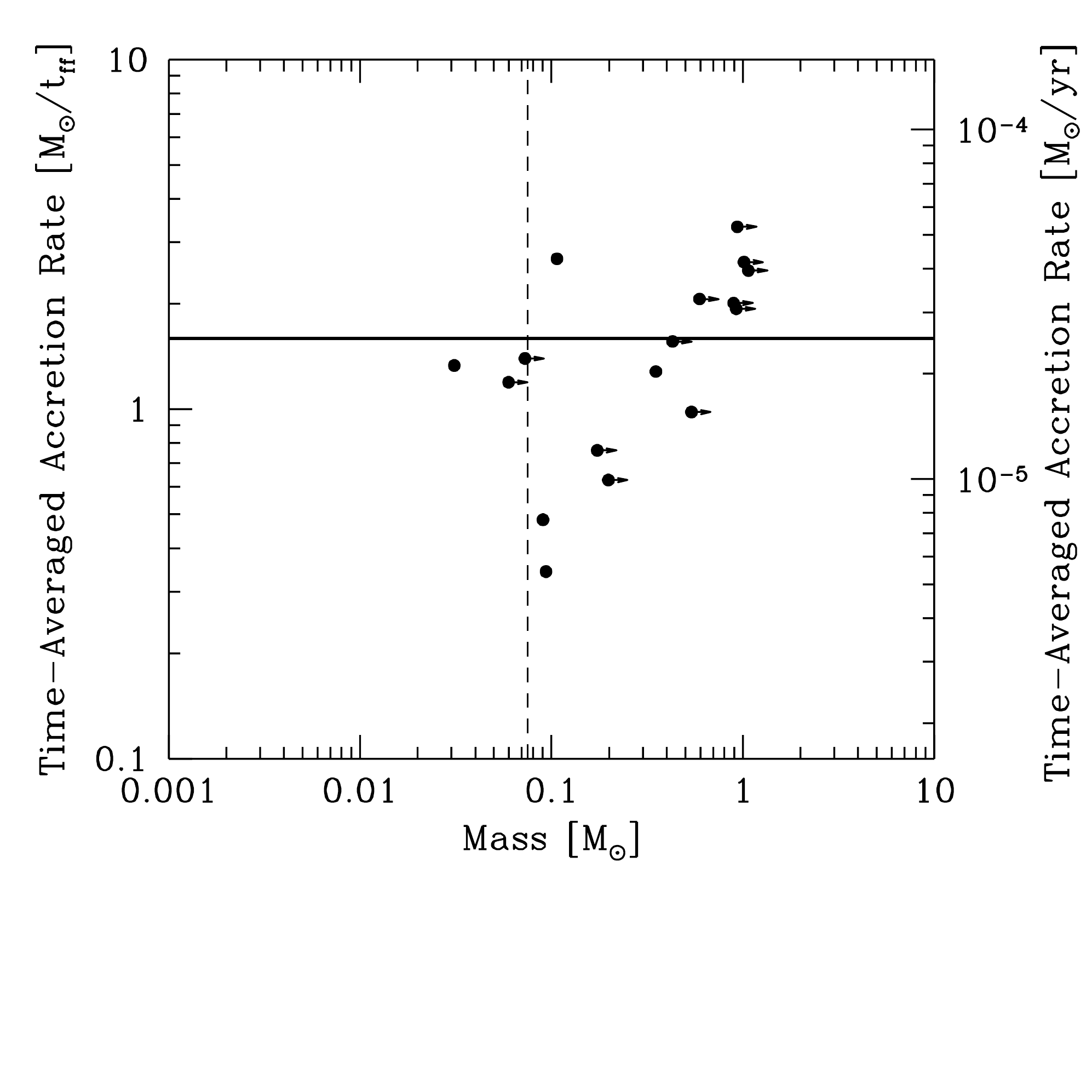}\vspace{-1.0cm}
\caption{The time-averaged accretion rates of the objects versus their final masses from the BBB2003 RT0.5 (top) and BB2005 RT0.5 (bottom) calculations.  The accretion rates are calculated as the final mass of an object divided by the time between its formation and the termination of its accretion or the end of the calculation.  The mean accretion rates for the two calculations are $1.5\times 10^{-5}$ M$_\odot$~yr$^{-1}$ and $2.5\times 10^{-5}$ M$_\odot$~yr$^{-1}$, respectively.  These values are statistically indistinguishable from each other given the small numbers of objects and the fact the the dispersions of the accretion rates are larger than the difference between the two calculations.  The horizontal solid lines give the mean accretion rates. Objects still accreting when the calculations were stopped are denoted with horizontal arrows.  The accretion rates are given in M$_\odot/t_{\rm ff}$ on the left-hand axes and M$_\odot$/yr on the right-hand axes. The vertical dashed line marks the star/brown dwarf boundary.
}
\label{accrate}
\end{figure}

Evaluating this equation is not straightforward because of the dependence on the protostellar luminosity which depends on the mass, radius and accretion rate of the protostar.  However, because the dependence on the protostellar luminosity is quite weak, we can put in estimated numbers to check that it gives a realistic characteristic mass.  The luminosity of an accreting protostar is
\begin{equation}
\label{accretionluminosity}
L_* \sim \frac{GM_* \dot{M_*}}{R_*},
\end{equation}
where $M_*$ and $R_*$ are the mass and radius of the protostar, respectively.
In the radiation hydrodynamics calculations presented here, the mean accretion rate of the objects is approximately the same for both calculations: $1.5\times 10^{-5}$ M$_\odot$~yr$^{-1}$ for BBB2003 RT0.5 and $2.5\times 10^{-5}$ M$_\odot$~yr$^{-1}$ for BB2005 RT0.5 (Figure \ref {accrate}).  These rates are calculated as the mean of the time-averaged accretion rates of all the objects in a calculation, where the time-averaged accretion rate is the mass of the object at the end of the calculation divided by the time between its formation and the end of its accretion or the end of the calculation, which ever comes first.  Assuming the radiative feedback typically comes from a 0.1 M$_\odot$ protostar of radius 2~R$_\odot$ accreting at a rate of $1\times 10^{-5}$ M$_\odot$~yr$^{-1}$ gives a luminosity of 150 L$_\odot$.  The effective Jeans mass for the BBB2003-type initial conditions with an initial cloud density of $1.2\times 10^{-19}$~g~cm$^{-3}$ can then be written
\begin{equation}
M_{\rm eff} \approx 0.5 \left( \frac{\rho}{1.2\times 10^{-19}~{\rm g~cm}^{-3}} \right)^{-\frac{1}{5}} \left( \frac{L_*}{150~{\rm L}_\odot} \right)^\frac{3}{10}   {\rm M}_\odot.
\end{equation}
This is a reasonable value for the characteristic mass of the IMF.  

To try and proceed beyond this rough estimate, we need to make further simplifying assumptions about the luminosity of the protostars.  The protostellar accretion rates may be expected to scale as  $\sim c_{\rm s}^3/G$.  However, even for an isothermal, non-singular spherical cloud the accretion rate usually begins much larger than $c_{\rm s}^3/G$ and declines with time \citep{FosChe1993}.  Here the situation is even more complex because the sound speed is that of the {\em internally heated} cloud which is a function of the radius from the protostar rather than simply the initial temperature of the cloud.  On the other hand, the timescale for the accretion of the envelope must still be $\sim \lambda_{\rm eff}/c_{\rm s}$ where the relevant sound speed is that at distance $\lambda_{\rm eff}$ from the protostar.  Therefore, we can write
\begin{equation}
\label{mdot}
\dot{M} \sim \frac{M_{\rm eff} c_{\rm s}}{\lambda_{\rm eff}} = \frac{4\pi \rho \lambda_{\rm eff}^2}{3} \left(\frac{\cal{R}}{\mu} T\right)^{1/2}.
\end{equation}
Again using equation \ref{temperature} to eliminate $T$ and then using equation \ref{accretionluminosity} to solve for $L_*$ we obtain
\begin{equation}
L_* = \left( \frac{4 \pi G M_*}{3 R_*}\right)^{8/7} \rho^{8/7} \lambda_{\rm eff}^2 \left(\frac{\cal{R}}{\mu}\right)^{4/7} \left( 4 \pi \sigma \right)^{-1/7}.
\end{equation}
Inserting this into equation \ref{generalradius} we get
\begin{equation}
\lambda_{\rm eff} \propto \rho^{-5/14} \left( \frac{M_*}{R_*} \right)^{1/7},
\end{equation}
which finally gives
\begin{equation}
\label{finalmass}
M_{\rm eff} \propto \left( \frac{M_*}{R_*} \right)^{3/7} \rho^{-1/14}.
\end{equation}
The interesting thing about this equation for the effective Jeans mass is that the dependence on the initial density of the cloud has been almost eliminated and all that remains is a dependence on the protostellar mass to radius ratio.  This result should be treated with caution, simply because of the number of simplifying assumptions that have been involved.  Furthermore, it is quite likely that the mass to radius ratio of the protostar may itself depend on the mass accretion rate in which case equation \ref{finalmass} still contains a dependence on the initial density of the molecular cloud via equation \ref{mdot}.  However, the main point to take away from this discussion is that in both equations \ref{jeansmass} and \ref{finalmass} the effective Jeans mass is at most only weakly dependent on the initial density of the molecular cloud which is consistent with our numerical results.

Finally, we note that we have assumed in the above discussion that the protostar's luminosity provides the dominant contribution to the temperature at distances $\lsim \lambda_{\rm eff}$ from the protostar, as opposed to the background temperature.  If this is not the case (i.e., essentially the protostellar luminosity is too weak to modify the Jeans length and mass) the relevant length scales and masses for collapse revert to the usual equations \ref{stdjeanslength} and \ref{stdjeansmass}.  This may occur for very low densities and/or high temperature clouds.  We have also assumed that the mean density of the cloud is low enough that the cloud is optically thin on length scales $\gg \lambda_{\rm eff}$ to the infrared radiation from the protostar.  If the density is high enough that the radiative feedback is diffusive  then using equations \ref{escape} and \ref{massradius}, but replacing equation \ref{temperature} with the diffusion approximation
\begin{equation}
L_* = - 4 \pi r^2 \frac{16~\sigma}{3~\kappa \rho} T^3 \frac{{\rm d}T}{{\rm d}r},
\end{equation}
the effective Jeans mass can be shown to scale as
\begin{equation}
M_{\rm eff} \propto \rho^{-1/3} L_*^{1/3}.
\end{equation}
Thus, its dependence on density is somewhat stronger (though still weaker than the standard Jeans mass) but the luminosity dependence is similar.  We do not discuss this case further here because it applies to densities and/or temperatures that are very different to local star-forming regions (e.g. mean densities $\gsim 10^8$ cm$^{-3}$ and temperatures $\gsim 30$~K).

In summary, it appears from both the numerical simulations and the above analytic arguments that radiative feedback may act to force a given amount of gas in a dense molecular core to form a particular number of stars, with only a very weak dependency of the number of stars on the density of the dense core.  This process can be seen in action in the BBB2003 RT0.5 and BB2005 RT0.5 calculations where the main dense cores with masses $\approx 10$~M$_\odot$ in both simulations each produce $\approx 10$ objects.  The characteristic length scale between protostars, $\lambda_{\rm eff}$, decreases with increasing density in such a way as to leave the characteristic mass only weakly dependent on the cloud's initial density.  We propose that this effective Jeans mass is responsible for setting the characteristic mass of the IMF, and that this may explain the apparent universality of the IMF observed in the local Universe.

Needless to say, the above analytic explanation of how radiative feedback diminishes the dependence of the characteristic mass of the IMF on the initial Jeans mass in the molecular cloud is a gross oversimplification.  In the calculations, and in reality, molecular clouds have highly inhomogenous densities, the protostellar accretion rates are variable both spatially and with time, and a protostar's luminosity depends on its accretion rate, mass, radius, and on the details of how discs funnel matter onto the protostar.  However, the above discussion does provide a framework within which we can begin to understand the role of radiative feedback in setting the characteristic mass.  Larger radiation hydrodynamical calculations that produce many more stars and brown dwarfs to provide better statistics will be needed to test whether or not the characteristic mass of the IMF does display the above weak dependencies on the initial conditions.

\section{Conclusions}
\label{conclusions}

We have repeated the previous hydrodynamical star cluster formation simulations of \cite{BatBonBro2003} and \cite{BatBon2005}, but using a realistic gas equation of state and radiative 
transfer in the flux-limited diffusion approximation rather than the original barotropic equation of
state.  We find that radiative feedback, even from low-mass protostars, has an enormous impact on the
star formation process.  Our conclusions are as follows.

\begin{enumerate}
\item Whereas star formation in the barotropic calculations continued unabated in each dense core until the simulations were stopped, radiative feedback from newly formed protostars strongly suppresses the production of new objects in low-mass ($\approx 10$~M$_\odot$) dense molecular cores after roughly one local dynamical time.
\item Radiative feedback inhibits the fragmentation of massive circumstellar discs surrounding newly-formed protostars and any dense filamentary gas in their vicinity.  This effect, together with (i), decreases the numbers of objects formed in the calculations by a factor of $\approx 4$ compared with the barotropic calculations.  However, it does not stop the formation of binary and multiple systems.  Even binaries with separations of only a few AU exist at the end of the calculations.  The components of these systems are typically widely separated when they form, but evolve to close systems via a combination of dynamical interactions, gas accretion, and interactions with discs.
\item The decrease in the fragmentation of discs and dense gas near existing protostars results in many fewer dynamical ejections than in the barotropic calculations.  Since dynamical ejections are responsible for the formation of brown dwarfs and low-mass stars in these calculations, the radiation hydrodynamical simulations produce many fewer brown dwarfs than in the barotropic calculations.  This results in a ratio of stars to brown dwarfs of $\approx 5$:1 that is in much closer agreement with observations than the barotropic simulations which produced more brown dwarfs than stars.
\item Whereas the characteristic stellar mass was found to scale linearly with the initial Jeans of the clouds in the barotropic calculations, with radiative feedback the two IMFs are indistinguishable.  We propose that the reason there is little observed variation of the IMF in the local Universe is because the star formation process self-regulates itself via radiative feedback.  Based on the numerical results, we present an analytic argument for how a characteristic mass based on radiative feedback from low-mass protostars might be derived that scales very weakly with the initial conditions in molecular clouds.  For example, assuming grey radiative transfer, we obtain a characteristic mass that scales as $M_{\rm eff} \propto \rho^{-1/5} L_*^{3/10}$, where $\rho$ is the density of the cloud and $L_*$ is the typical  protostellar luminosity.
\item Finally, we note that due to the sink particle approximation used in the radiation hydrodynamical calculations presented here, the protostellar luminosity is underestimated.  The intrinsic protostellar luminosity and a substantial fraction of the accretion luminosity is neglected.  Thus, the dramatic effects of radiative transfer presented here are actually lower-limits.  We investigate the degree to which this may affect our results by performing one of the two calculations with different sink particle parameters (accretion radii of 0.5 and 5.0 AU).  Future calculations should attempt to include the additional radiative feedback from within the unresolved regions surrounding the protostars.
\end{enumerate}

\section*{Acknowledgments}

MRB is grateful for useful discussions with Jim Pringle, Christophe Pinte, Sergei Nayakshin, and Tim Harries.  
The computations reported here were performed using the UK Astrophysical Fluids Facility (UKAFF) and the University of Exeter Supercomputer. 
MRB is also grateful for the support of a Philip Leverhulme Prize and a EURYI Award. 
This work, conducted as part of the award ÒThe formation of stars and planets: Radiation hydrodynamical and magnetohydrodynamical simulationsÓ made under the European Heads of Research Councils and European Science Foundation EURYI (European Young Investigator) Awards scheme, was supported by funds from the Participating Organisations of EURYI and the EC Sixth Framework Programme. 

\bibliography{mbate}

\end{document}